\documentclass[useAMS,usenatbib]{mnras}
\usepackage{epsfig}
\usepackage{color} 
\usepackage{natbib}
\usepackage{graphicx}
\usepackage{booktabs}
\usepackage{amssymb}
\usepackage{array}
\def\ew{$W_{2796}$}
\def\ewfeii{$W_{2600}$}
\def\ewmgi{$W_{2852}$}
\def\hi{H~{\sc i}~} 
\def\nhi{$N${\sc (H~i)}}

\def\ls{L$_{\star}$~}
\def\lsoii{L$^{\star}_{[\rm O~{\sc II}]}$}

\def\sloii{$\Sigma$$_{\rm [O~II]}$}
\def\ssfr{$\Sigma$$_{\rm SFR}$}
\def\loii{L$_{[\rm O~II]}$}

\def\mgi{Mg~{\sc i}~} 
\def\mgii{Mg~{\sc ii}~} 
\def\caii{Ca~{\sc ii}~}

\newcommand{\rratio}{\rm {\ensuremath{\mathcal{R} \equiv}}  {\ensuremath{W_{\rm 2600}}}/{\ensuremath{W_{\rm 2796}}}~}
\newcommand{\rrr}{\rm {\ensuremath{\mathcal{R}}}} 
\def\feii{Fe~{\sc ii}~} 
 
\def\feiia{Fe~{\sc ii}$\lambda$2600~} 

\def\oiiab{[O~{\sc ii}]$\lambda\lambda$3727,3729}  
  
\def\oii{[O~{\sc ii}]}

\def\o3o2{[O~{\sc iii}]/[O~{\sc ii}]}
\def\zabs{$z_{\rm abs}$} 
\def\zem{$z_{\rm em}$~} 
 
\def\lya{Ly$\alpha$~}

\def\kms{km\ s$^{-1}$} 
\def\aj{{AJ}}%
\def\apj{{ApJ}}%
\def\apjl{{ApJ}}%
\def\apjs{{ApJS}}%
\def\aap{{A\&A}}%
\def\mnras{{MNRAS}}%
\title[\oii\ emission associated with \mgii absorbers]{Average \oii\  nebular emission  associated with \mgii absorbers: Dependence  on \feii absorption} \author[Joshi, R. et al.]{Ravi Joshi$^{1}$\thanks{E-mail: rvjoshirv@gmail.com(RJ)}, Raghunathan Srianand $^{2}$, Patrick  Petitjean$^{3}$  and  Pasquier Noterdaeme$^{3}$  \\
$^{1}$Kavli Institute for Astronomy and Astrophysics, Peking University, Beijing 100871, China \\
$^{2}$Inter-University Centre for Astronomy and Astrophysics, Post Bag 4, Ganeshkhind, Pune 411007, India \\
$^{3}$UPMC-CNRS, UMR7095, Institut d'Astrophysique de Paris, F$-$75014 Paris, France \\
}

\begin{document}
\date{Accepted ---. Received ---; in original form ---}

\pagerange{\pageref{firstpage}--\pageref{lastpage}} \pubyear{2018}

\maketitle

\label{firstpage}
\begin{abstract}

We investigate the effect of \feii equivalent width (\ewfeii) and
fibre size on the average luminosity of \oiiab\ nebular emission
associated with \mgii absorbers (at $0.55 \le z \le 1.3$) in the
composite spectra of quasars obtained with 3 and 2 arcsec fibres in
the Sloan Digital Sky Survey. We confirm the presence of strong
correlations between \oii\ luminosity (\loii) and equivalent width
(\ew) and redshift of \mgii absorbers. However, we show \loii\ and
average luminosity surface density suffers from fibre size effects.
More importantly, for a given fibre size the average \loii\ strongly
depends on the equivalent width of \feii absorption lines and found to
be higher for \mgii absorbers with \rratio $\ge 0.5$. In fact, we show
the observed strong correlations of \loii\ with \ew\ and $z$ of \mgii
absorbers are mainly driven by such systems. Direct \oii\ detections
also confirm the link between \loii\ and \rrr. Therefore, one has to
pay attention to the fibre losses and dependence of redshift evolution
of \mgii absorbers on \ewfeii\ before using them as a luminosity
unbiased probe of global star formation rate density. We show that the
\oii\ nebular emission detected in the stacked spectrum is not
dominated by few direct detections (i.e., detections $\ge 3 \sigma$
significant level). On an average the systems with \rrr\ $\ge 0.5$ and
\ew\ $\ge 2$\AA\ are more reddened, showing colour excess E($B-V$)
$\sim$ 0.02, with respect to the systems with \rrr\ $< 0.5$ and most
likely traces the high \hi column density systems.

\end{abstract}


\section{Introduction}
\label{sec:intro_mgiifeii}

Understanding the surface star-formation rate (\ssfr) and its redshift
dependence associated with quasar absorbers is of utmost importance
for measuring the star-formation rate density (SFRD) as a function of
cosmic time in a luminosity independent way
\citep{Wolfe2003ApJ...593..235W,Srianand2005MNRAS.362..549S,Rahmani2010MNRAS.409L..59R,Menard2011MNRAS.417..801M}
and to probe the Kennicutt-Schmidt law in low metallicity gas at high
redshifts \citep{Rahmani2010MNRAS.409L..59R}. Evidence for large scale
outflows are seen in nearly all star-forming galaxies in the local
Universe and out to z $\sim$ 6
\citep{Pettini2001ApJ...554..981P,Shapley2003ApJ...588...65S,Steidel2010ApJ...717..289S,Martin2012ApJ...760..127M,
  Newman2012ApJ...761...43N,Lundgren2012ApJ...760...49L,Zhu2015ApJ...815...48Z}.
Wind detection rates in galaxies at intermediate and low redshifts are
found to be depending on the galaxy orientation with the outflow
geometry being consistent with a bi-conical flow
\citep{Martin2012ApJ...760..127M}. These winds are thought to be
responsible for enriching the intergalactic medium (IGM) and
circumgalactic medium (CGM) around galaxies
\citep{Schaye2001ApJ...559L...1S,Simcoe2012Natur.492...79S}. At high
redshifts, for isolated galaxies, gas inflowing rate is found to be
roughly comparable to the sum of the star formation rate and the
outflowing rate
\citep{Erb2008ApJ...674..151E,Seko2016ApJ...833...53S}. \par

 At present, the best way to probe the low density outflowing gas is
 to study the metal absorption lines they imprint in the spectra of
 background luminous sources (at small impact parameters; $\rho < 10$
 kpc) which trace the dynamic environment, i.e., gas inflows and
 outflows in the outskirt of galaxies over cosmic time-line
 \citep{Weisheit1978ApJ...219..829W,
   Lanzetta1992ApJ...391...48L,Mo1996ApJ...469..589M,
   Tinker2008ApJ...679.1218T,Chelouche2010ApJ...722.1821C,Bouche2012MNRAS.426..801B}.
 In the local Universe winds are ubiquitous in galaxies having
 \ssfr\ $ \ge 0.1\ \rm {M_\odot\ yr^{-1}\ kpc^{-2}}$
 \citep[][]{Heckman2001ASPC..240..345H,Heckman2002ASPC..254..292H}. If
 high \mgii equivalent width systems are associated with outflows, as
 suggested by \citet{Nestor2011MNRAS.412.1559N} and
 \citet{Bouche2012MNRAS.426..801B}, then the associated galaxy is
 expected to have high \ssfr. Therefore, understanding the connection
 between absorber properties and their associated galaxies (e.g.,
 equivalent width vs \ssfr) is vital for understanding various
 feedback processes that shape up the galaxy evolution.

Interestingly, the average \ssfr\ per absorber can be obtained using
spectral stacking exercise
\citep{Wild2007MNRAS.374..292W,Noterdaeme2010MNRAS.403..906N,
  Menard2011MNRAS.417..801M} by making suitable assumptions related to
fibre losses. Note that, in spectroscopic surveys using fibres (e.g.,
3 and 2 arcsec fibres used in SDSS-DR7 and DR12 surveys) one not only
integrate light from the background quasar but also from all
foreground galaxies that happen to fall within the fibre. However,
because of various practical reasons it is highly probable that only a
part of the line emitting region of foreground galaxy may come inside
the fibre and this leads to the fiber loss. This loss will also have
redshift dependence. At low redshifts $0.4 < z < 1.3$, in a stacking
experiment of 3461 \mgii and 345 \caii\ absorbers,
\citet{Wild2007MNRAS.374..292W} have detected the \oii\ nebular
emission and measured an average star formation rate (SFR) of
$0.11-0.14~ {\rm M_\odot\ yr^{-1}}$ and $0.11-0.48~ {\rm
  M_\odot\ yr^{-1}}$, respectively. Using the enlarged sample of \mgii
absorbers from SDSS-DR7, \citet{Noterdaeme2010MNRAS.403..906N} have
detected the average \oii\ luminosity of $\sim$ $1.4-5.1 \times
10^{40} \rm ~erg\ s^{-1}$ for strong \mgii\ absorbers (i.e., \ew\ $\ge
1$~\AA) at $0.5 < z < 0.7$. \citet{Menard2011MNRAS.417..801M} proposed
that \mgii absorbers recover the overall star formation history of the
universe and can be used as a new tool to probe the redshift evolution
of SFR in a luminosity independent manner.
\citet{Rahmani2010MNRAS.409L..59R} have stacked the damped-\lya
absorbers (DLA) with log \nhi $\ge 20.3$ to detect the average \lya
emission and set an upper limit on the contribution of DLA galaxies to
the cosmic SFRD of $\sim$ 0.13~ ${\rm M_\odot\ yr^{-1}\ Mpc^{-3}}$ at
$z \sim 3$ \citep[see
  also,][]{Noterdaeme2014A&A...566A..24N,Joshi2017MNRAS.465..701J}.
However, as mentioned above physical parameters derived from fibre
spectra suffer from effect of fibre losses and one has to be mindful
of this while interpreting these results
\citep{Lopez2012MNRAS.419.3553L, Joshi2017MNRAS.471.1910J}.

Fibre losses can in principle be considered as negligible for systems
with large \ew\ because of the well known anticorrelation between
impact parameter ($\rho$) and \ew\ \citep{Bergeron1986A&A...155L...8B,
  Steidel1995qal..conf..139S, Chen2010ApJ...714.1521C,
  Nielsen2013ApJ...776..114N}. \citet{Lopez2012MNRAS.419.3553L} have
shown that for the fibre size used in SDSS-II (3 arcsec), galaxies
associated with more than 90\% of the systems will fall inside the
fibre for \ew $> 3$~\AA\ (see their figure 4). However, large scatter
in \ew\ vs $\rho$ relationship could increase the fibre loss. A strong
correlation is seen between the \mgii equivalent width (at low
spectral resolution is a good proxy to the velocity spread along the
line of sight) and galaxy color where a stronger \mgii absorbers tend
to be present in the vicinity of star-forming galaxies and most likely
to be associated with the outflows \citep{Zibetti2007ApJ...658..161Z,
  Noterdaeme2010MNRAS.403..906N, Bordoloi2011ApJ...743...10B,
  Lan2014ApJ...795...31L,Nielsen2016ApJ...818..171N}. The absorbing
gas traced by \mgii absorbers appear to have a bimodality in azimuthal
angle distribution where the cool dense gas is preferred to lie near
major and minor axes of galaxies \citep{Bouche2012MNRAS.426..801B,
  Kacprzak2012ApJ...760L...7K}.

The \mgii absorbers with strong \feii\ absorption are likely to arise
from either very high metallicity sub-DLAs or high \nhi\ DLAs
\citep{Srianand1996ApJ...462..643S, Rao2006ApJ...636..610R}.
\citet{Rao2006ApJ...636..610R} have found that the detection rate of
DLAs in \mgii systems increases if one puts additional constrains
based on equivalent width ratios of Mg~{\sc ii}, \mgi and \feii
absorption. They detected DLAs with a success rate of $\sim$ 42 per
cent by selecting \mgii absorbers with strong \feii absorption (i.e.,
\rratio $\ge 0.5$) and \ewmgi\ $> 0.1$~\AA. In recent efforts to
detect cold gas in strong \mgii systems
\citet{Dutta2017MNRAS.465.4249D} have found a factor four times higher
detection rate of \hi 21-cm absorption in systems with \ewfeii\ $\ge
1$~\AA\ \citep[see also,][]{Gupta2012A&A...544A..21G}. Recently, we
have detected 198 \oii\ emitting galaxies associated with strong \mgii
absorbers in the spectra of SDSS \citep[][hereinafter refer as Paper
  1]{Joshi2017MNRAS.471.1910J}. We have found that the \mgii absorbers
detected in \oii\ nebular emission (with \loii\ $> 2 \times 10^{40}
\rm ~erg\ s^{-1}$) typically have \ew\ $\ge$ 1~\AA, \mgii doublet
ratio ($DR = W_{\rm Mg~II \lambda2796}/W_{\rm Mg~II \lambda2803}$)
close to unity and \rrr $\ge$ 0.5. Therefore, given the above facts,
naively one would expect a strong dependence of average
\oii\ luminosity of \mgii absorbers in the stacked spectra of systems
with different \rrr\ parameter ranges. \par

There are two main motivations behind this work: (i) to study the
dependence of average \loii\ on \rrr\ and (ii) to understand the
effect of fibre size on different observed correlations between
\loii\ and other parameters. While the former allows us to probe the
nature of SFR in potential DLA candidates the latter will allow us to
probe the gas distribution at different scale around the star forming
regions probed by different samples of \mgii\ absorbers.

\begin{figure}
  \centering
 \epsfig{figure=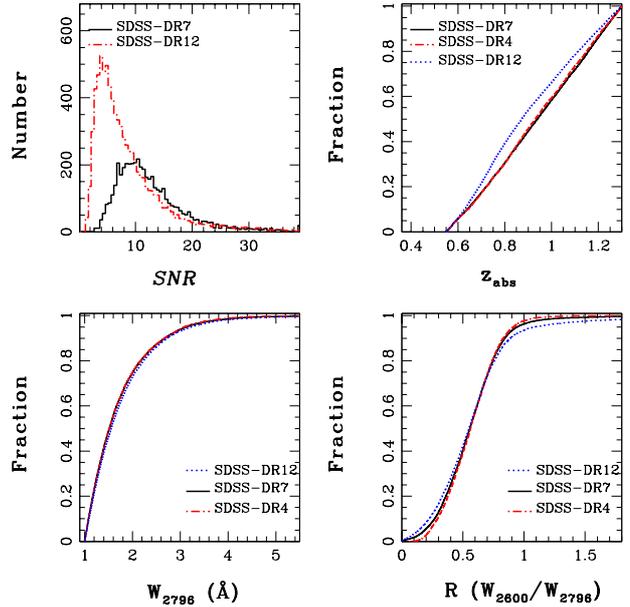,height=8.5cm,width=8.5cm,angle=0}
  \caption{\emph{Top left panel:} The signal-to-noise ratio
    distribution of quasar spectra around the expected position of
    \oii\ emission line in SDSS-DR7 and DR12 spectra. The cumulative
    distribution of the \zabs\ (\emph{top right}), \ew (\emph{bottom
      left}) and \rratio (\emph{bottom right}) of our \mgii systems
    from SDSS-DR7 and SDSS-DR4 \mgii systems used in
    \citet{Menard2011MNRAS.417..801M}.}
\label{fig:sample_comp}
 \end{figure}

\begin{figure*}
  \centering
 \epsfig{figure=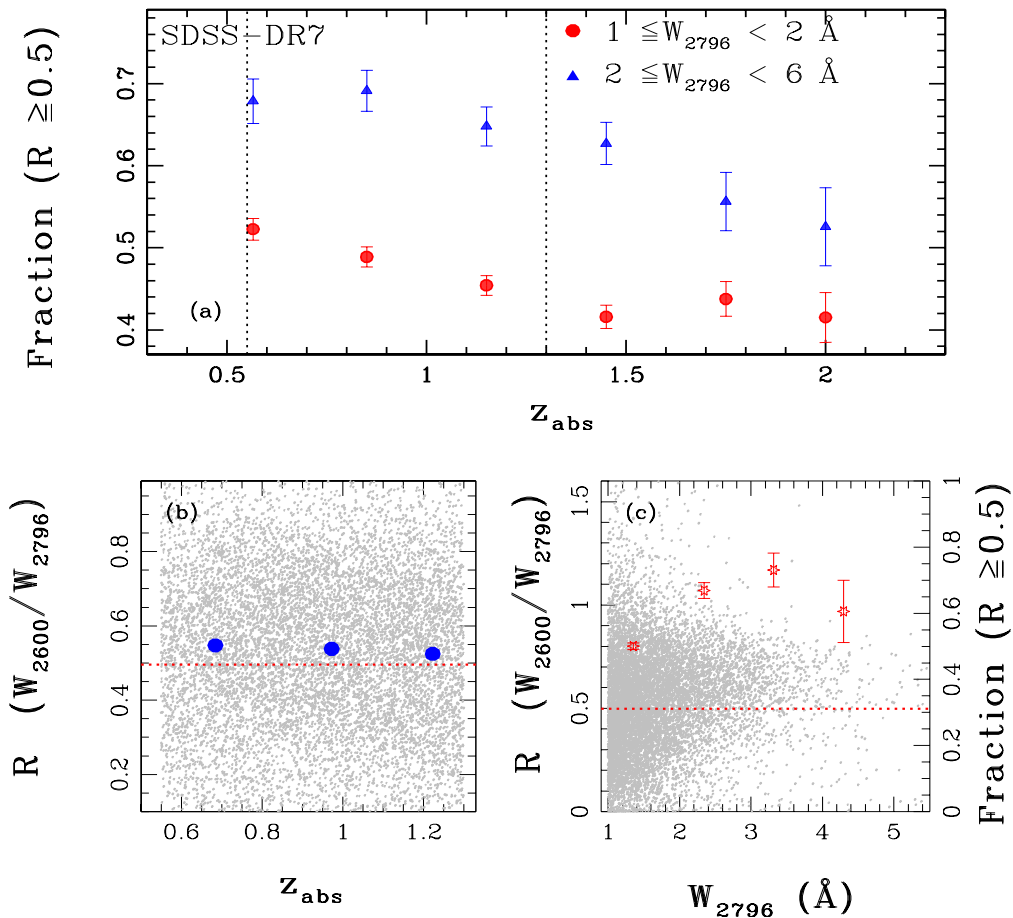,height=12.5cm,width=13.5cm,angle=0}
  \caption{\emph{Top panel:} The fraction of \mgii systems with $R \ge
    0.5$ versus $z$ for two different \ew\ ranges. Vertical dotted
    lines mark the redshift range of interest for this study.
    \emph{Bottom right panel:} The dependence of \rrr\ parameter on
    \ew\ for the \mgii systems. The corresponding fraction is shown as
    \emph{stars} in the right side ordinate. For this plot we consider
    only \mgii systems in the redshift range of our interest.
    \emph{Bottom left panel:} \rrr\ as a function of redshift for
    systems with \ew\ $\ge 1$~\AA. The solid blue circles show the
    median value of \rrr\ in different redshift bins. The results are
    for \mgii\ systems from SDSS-DR7 (see Fig~\ref{fig:fraction_dr12}
    in Appendix~\ref{sec:apd1} for the results based on SDSS-DR12).}
\label{fig:fraction}
 \end{figure*}

This article is organized as follows. Section 2 describes our sample
of absorbers used in this study. In Section 3, we present the spectral
staking analysis. In Section 4, we present various correlations seen
in the stacked spectra of quasars from SDSS-DR7 and DR12. Here, we
also discuss the dependence of \oii\ on \rrr\ parameter. Finally,
discussions and conclusions are presented in Section 5. Throughout, we
have assumed a flat cosmology with $H_0 =$ 70~\kms\ $\rm Mpc^{-1}$,
$\Omega_{\rm m} = 0.3$ and $\Omega_{\rm \Lambda} = 0.7$.

\section{Sample}
\label{sec:sample}

  We have constructed a sample of strong \mgii absorbers, defined as
  the ones with rest equivalent width \ew\ $\ge 1$~\AA\ (detected at
  $\ge 4 \sigma$), by using the compilation of \mgii systems from the
  \emph{expanded-version} of JHU-SDSS Metal Absorption Line
  Catalog{\footnote{\href{http://www.pha.jhu.edu/$\sim$gz323/Site/Download\_Absorber\_Catalog\_files/fits/}
      {http://www.pha.jhu.edu/$\sim$gz323/Site/}}
    \citep[][]{Zhu2013ApJ...770..130Z}, compiled from the SDSS$-$DR7
    and SDSS$-$DR12. For a fully saturated line this equivalent width
    threshold corresponds to a velocity spread of $\sim 107$~\kms. We
    select only systems with velocity offset of $>$ 5000~\kms\ with
    respect to the quasar emission redshift and avoid sightlines
    having broad absorption lines produced by quasar outflows (i.e.,
    BALQSOs). To investigate the dependence of the average
    \oii\ luminosity in the stacked spectra on \rrr\ parameter we
    ensured that the \feiia line falls in the wavelength range of
    higher completeness at $\lambda >$ 4000~\AA, i.e., $z \ge 0.55$
    and redward of \lya emission of the quasar. In addition, we
    restrict ourselves to $z \le 1.3$ (i.e. $\lambda \lesssim
    8500$~\AA\ for \oiiab\ line) to avoid the \oii\ emission being
    contaminated by most crowded telluric lines. For the above
    redshift range, i.e., $ 0.55 \le z \le 1.3$, the fibres of 3 and 2
    arcsec diameter used for the SDSS-DR7 and DR12 observations
    project an angular size of $\sim 9.6-12.5$ kpc and $\sim 6.4-8.4$
    kpc, respectively, in the sky. Our final sample consists of 10,083
    and 12,116 \mgii systems from SDSS-DR7 and DR12, respectively.
    \par

    We note that our sample of \mgii systems based on SDSS-DR7 is
    similar to the SDSS-DR4 sample used by
    \citet{Menard2011MNRAS.417..801M} in terms of \zabs, \ew\ and
    \rratio (see also, Fig.~\ref{fig:sample_comp}). A two-sided
    Kolmogorov-Smirnov test (KS-test) finds no difference between the
    two sub-samples based on \zabs, \ew\ and \rrr\ parameters with a
    null probability of being drawn from the same parent distribution
    to be $P_{KS} =$ 0.8, 0.8 and 0.7, respectively. The median
    redshift probed in SDSS-DR7 and DR4 sample is \zabs\ $\sim$ 0.9.
    However, SDSS-DR7 has twice the number of \mgii systems in the
    sample of \citet{Menard2011MNRAS.417..801M}. The \mgii systems
    detected in SDSS-DR12 also show a similar \ew\ distribution as
    seen in SDSS-DR7 and DR4. However, the absorption redshifts
    (\zabs) of \mgii systems in SDSS-DR12 are found to be slightly
    lower with a median \zabs\ $\sim$ 0.8. One can note that the
    average $SNR$, measured around the expected \oii\ nebular
    emission, for the SDSS-DR7 spectra is slightly higher than the
    SDSS-DR12 spectra (top left panel in Fig.~\ref{fig:sample_comp}).
    The mean (respectively, median) \rrr\ parameter of \mgii systems
    from SDSS-DR4, DR7 and DR12 is found to be $\sim$ 0.57 (0.56),
    $\sim$ 0.56 (0.56), $\sim$ 0.55 (0.54), respectively.

    In panel (a) of Fig.~\ref{fig:fraction} we show the evolution of
    fraction of strong \mgii absorbers with \rrr\ $\ge 0.5$ as a
    function of redshift. It is clear from the figure that over the
    redshift range of our interest (i.e. $0.55 \le z \le 1.3$
    identified with vertical dotted lines in Fig.~\ref{fig:fraction})
    the fraction of absorbers with \rrr $\ge 0.5$ in the range $\rm
    1~\AA\ \le$ \ew\ $< 2$~\AA\ decreases from 53\% to 44\%. However,
    this fraction remains nearly constant at $\sim$ 70\% for the
    systems with $\rm 2~\AA\ \le$ \ew\ $\le 6$~\AA\ over the same
    redshift range (i.e., region between the two dotted lines in
    Fig.~\ref{fig:fraction}). Even these strong systems show
    decreasing trend when we consider the full observed $z$ range.
    \par

    \citet{Dey2015MNRAS.451.1806D} have found the median \rrr\ values
    to decrease with increasing $z$ for \mgii systems detected in
    SDSS-DR7. We see the same trend with \mgii\ systems detected in
    SDSS-DR12 for the full sample (see Fig.~\ref{fig:fraction_dr12} in
    Appendix~\ref{sec:apd1}). Based on their fit (see their figure 2)
    we except {\rm {\ensuremath{\mathcal{R}}}} to decrease by 0.04
    between redshift 0.5 and 1.3. For systems in the restricted
    redshift range of our interest (i.e., $0.55 \le z \le 1.3$) we do
    see a similar evolution of {\rm {\ensuremath{\mathcal{R}}}} with
    redshift where {\rm {\ensuremath{\mathcal{R}}}} decreases by
    $\sim$0.03 (see panel b of Fig.~\ref{fig:fraction}). They
    interpreted this evolution to be due to evolution in the
    metallicity ratio $\rm [Fe/Mg]$, most probably caused by the
    cosmic evolution in the SNIa rates. In panel (c) of
    Fig.~\ref{fig:fraction} we plot \rrr\ as a function of \ew\ for
    all the systems in the $z$ range of our interest. The dependence
    of \rrr\ on \ew\ is apparent as the fraction of systems with
    \rrr\ $\ge 0.5$ increases from $\sim$ 50\% to $\sim$ 80\% for the
    \ew\ range from 1~\AA\ to 3~\AA.

\begin{figure*}
  \centering
 \epsfig{figure=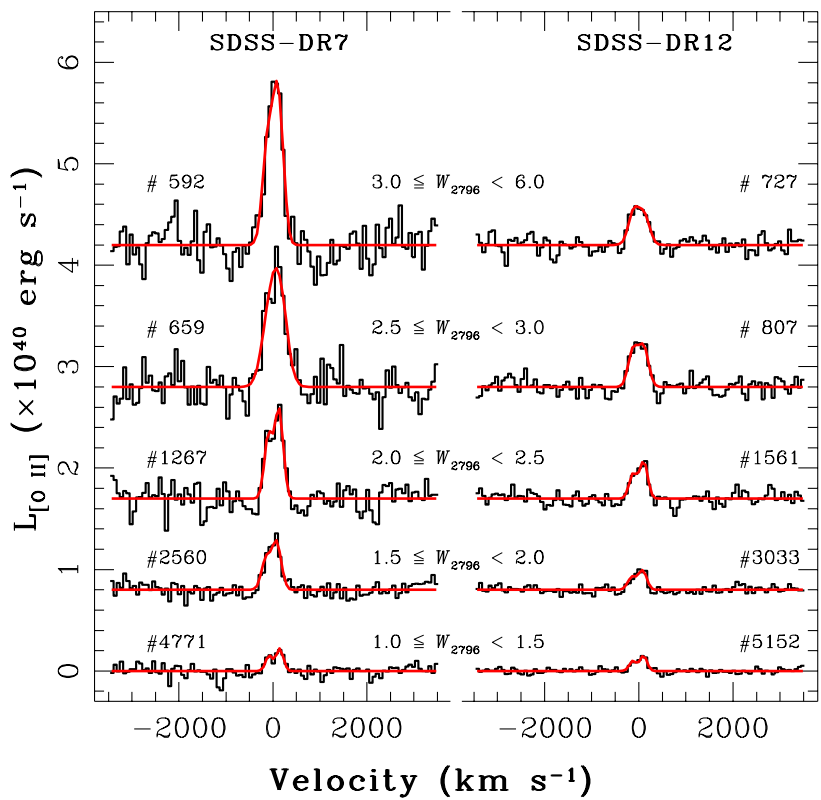,height=12.0cm,width=12.0cm,angle=0}
  \caption{The \oii\ emission line profile seen in SDSS-DR7
    (\emph{left}) and DR12 (\emph{right}) stacked spectra for various
    \ew\ bins. The \emph{solid} line shows the best fit double
    Gaussian to the data. The continuum-subtracted spectra are shifted
    by a constant offset in luminosity for display purpose.}
\label{fig:profile}
 \end{figure*}

\begin{figure*}
  \centering
 \epsfig{figure=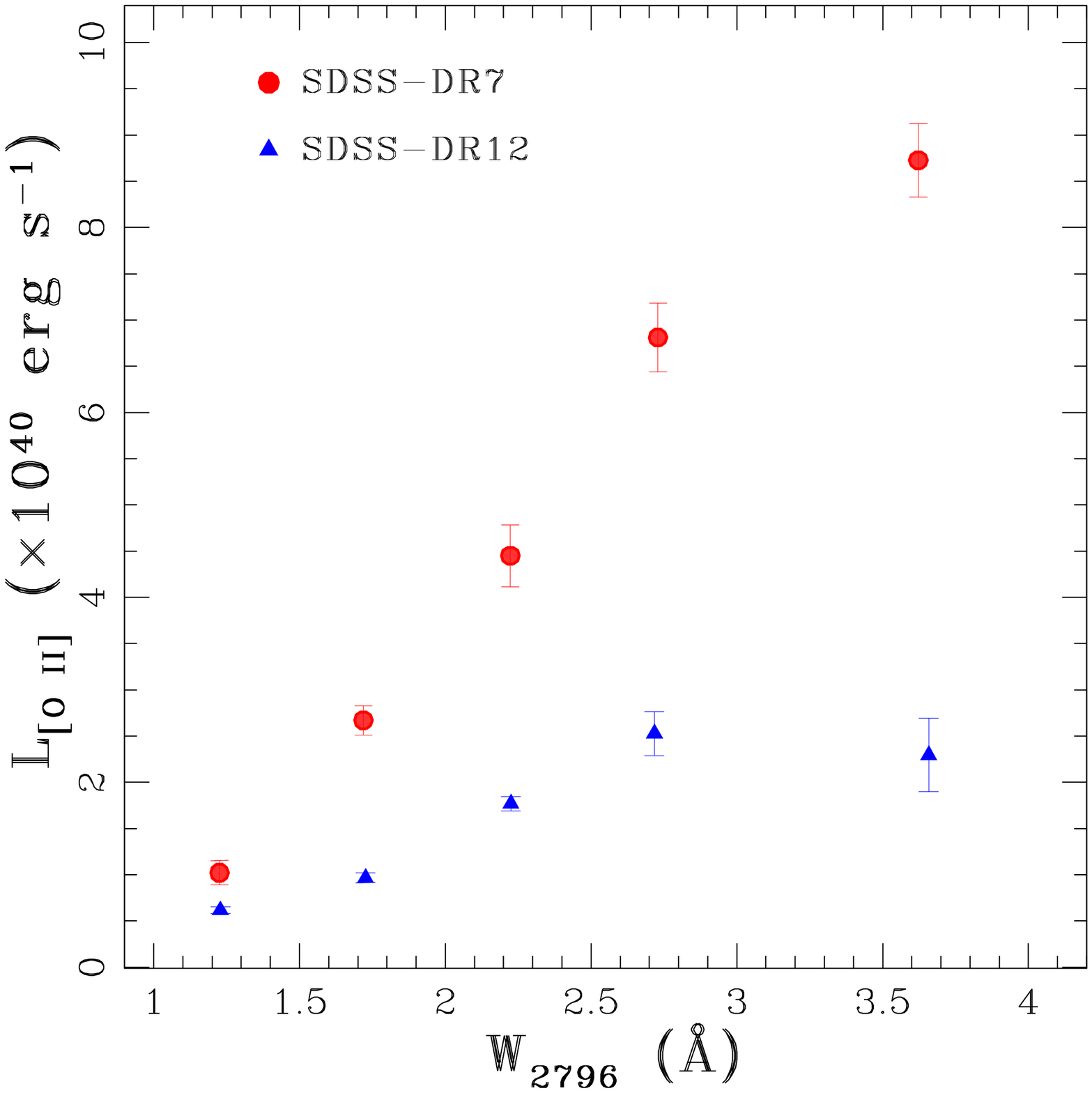,height=8.0cm,width=8.0cm,angle=0}
 \epsfig{figure=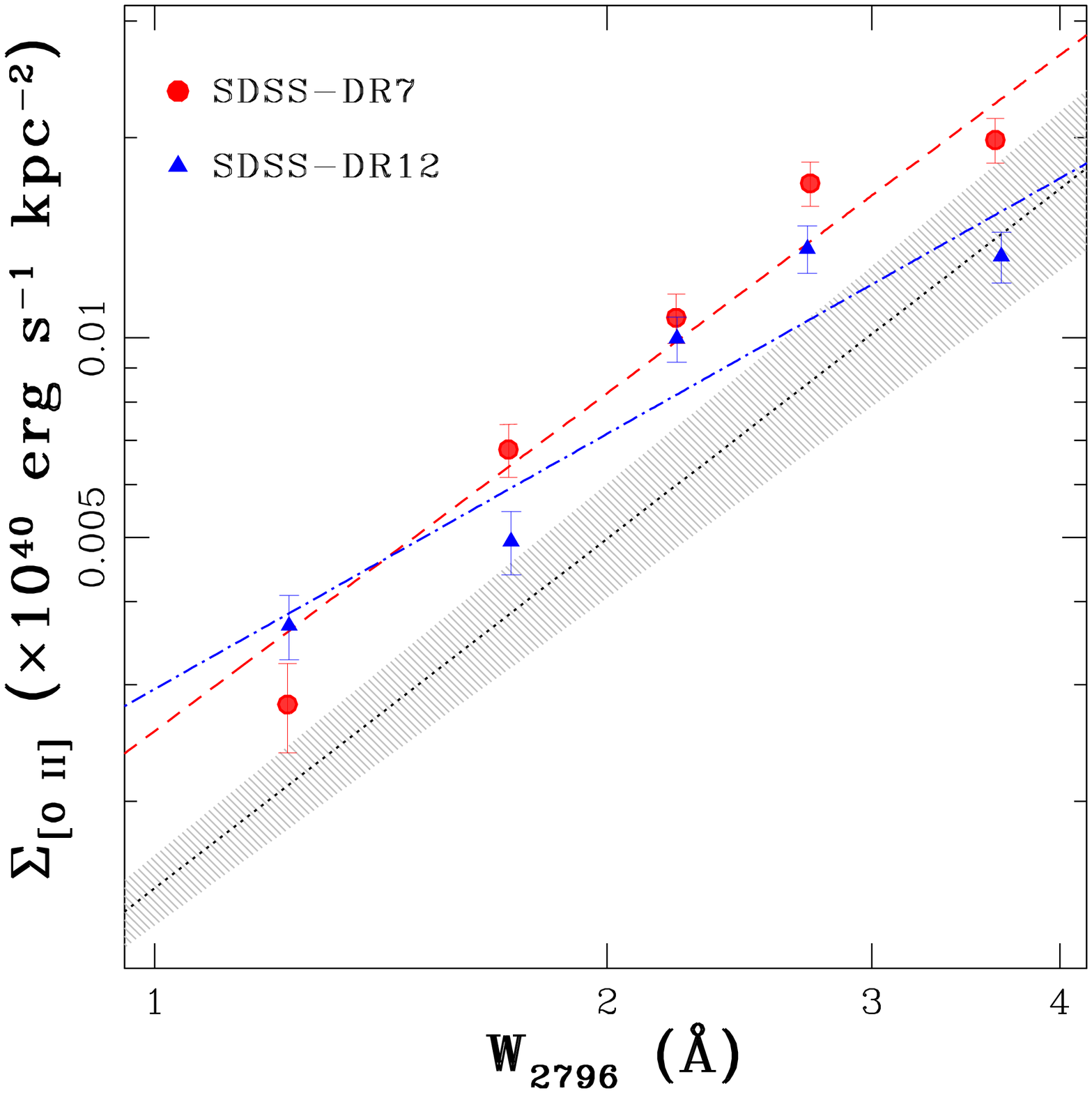,height=8cm,width=8cm,angle=0}
  \caption{\emph{Left panel:} The \oii\ luminosity of \mgii absorbers
    for various \ew\ bins for SDSS-DR7 (\emph{red circles}) and DR12
    (\emph{blue triangles}). \emph{Right panel:} The \oii\ luminosity
    surface density (\sloii) as a function of \ew\ for SDSS-DR7
    (\emph{red circles}) and DR12 (\emph{blue triangles}). The
    \emph{dotted line } shows the best fit for \sloii\ versus
    \ew\ from \citet{Menard2011MNRAS.417..801M} while shaded region
    show the 1$\sigma$ uncertainty (see Appendix~\ref{sec:apd2} for
    more discussion on this difference). The \emph{dashed} and
    \emph{dot-dashed} line are the best-fitting power law for our
    measurements based on SDSS-DR7 and DR12 data set.}
\label{fig:loiivsew}
 \end{figure*}

\begin{figure}
  \centering
 \epsfig{figure=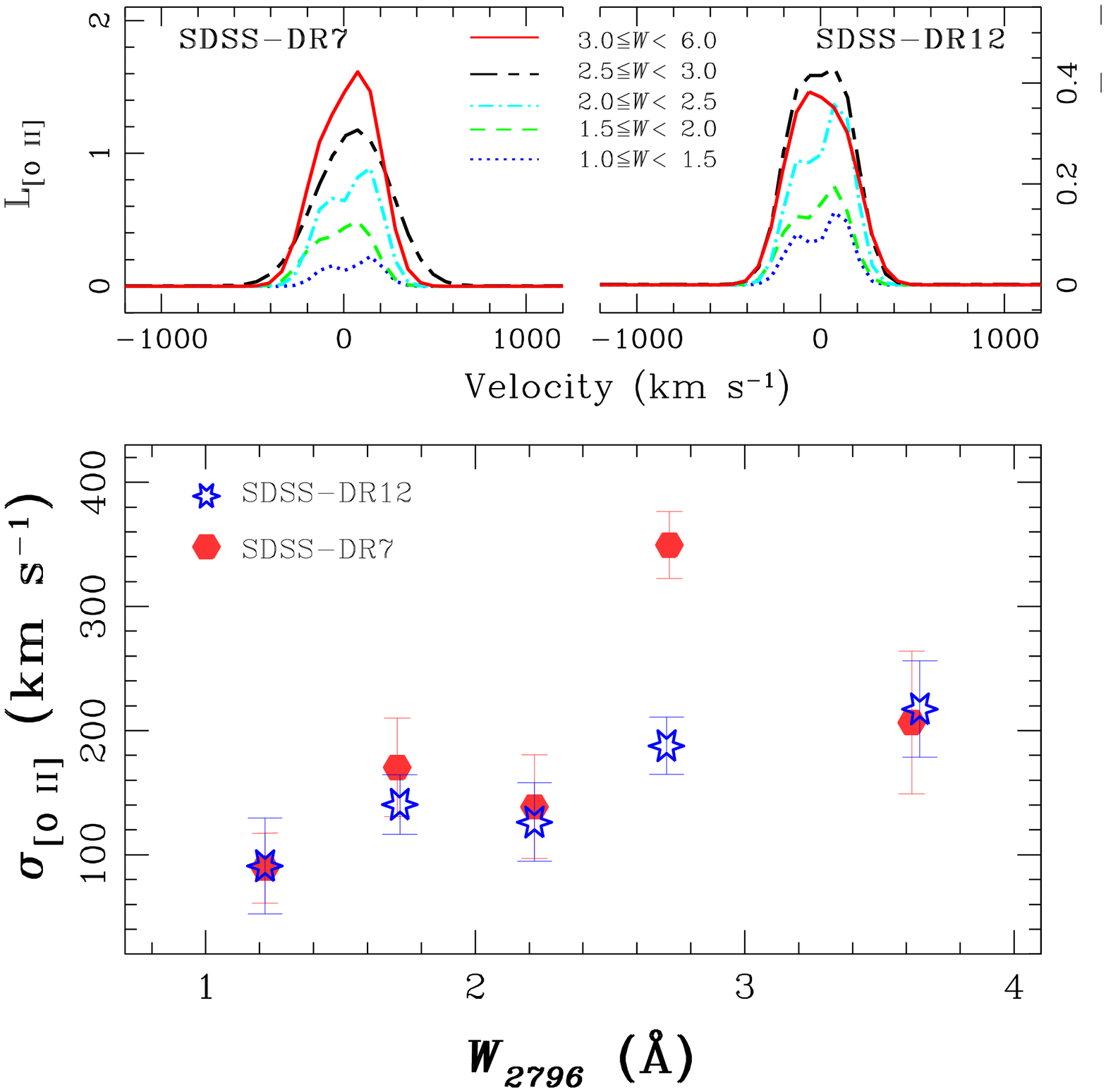,height=8.5cm,width=8.5cm,angle=0,bbllx=20bp,bblly=145bp,bburx=589bp,bbury=712bp,clip=true}
  \caption{\emph{Lower panel:} The distribution of $\sigma_{\rm [O
        II]}$ as a function of \ew\ for SDSS-DR7 (\emph{diamonds}) and
    SDSS-DR12 (\emph{stars}). \emph{Upper panel:} shows the \oii\ line
    profile for various \ew\ bins for SDSS-DR7 and DR12.}
\label{fig:sigma_delv}
 \end{figure}

\section{Analysis}
\label{sec:analysis}

To detect the \oii\ nebular emission from \mgii absorbers we have
constructed composite spectra, using continuum subtracted spectra and
median statistics. For this, we have shifted the individual spectrum
to the rest-frame of the \mgii absorber by conserving the flux and
rebinning on to a uniform rest wavelength grid as the original data
\citep{Bolton2012AJ....144..144B}. We modeled the local continuum by a
low order (typically a third order) polynomial fit, within the
proximity of \oiiab\ line (i.e., rest wavelength range of 3700 $-$
3750~\AA). While stacking, we have masked the absorption line features
originating from systems at other redshifts as well as the sky
emission lines in the spectrum. The 1$\sigma$ flux uncertainty in each
pixel in the stacked spectrum is estimated from the central interval
encompassing 68\% of the flux distribution of the corresponding pixel
as in \citet{Joshi2017MNRAS.465..701J}.

Recall that the physical area corresponding to a given angular
aperture varies with redshift. As suggested by
\citet{Menard2011MNRAS.417..801M}, to account for the fibre effects,
we also generate the stacked spectrum of \oii\ luminosity surface
density (\sloii). To estimate the \sloii, we convert each spectrum
into luminosity units, at the redshift of the absorber and divide by
the projected surface area of the fibre at the absorber redshift
before co-adding them to get the composite spectrum that we will call
\sloii.

As expected we detect \oii\ emission in most of our composite spectra.
In each composite spectrum we model the observed \oii\ emission line
with a double Gaussian profile with a tied linewidth but freely
varying the line ratio in a range of 3.4-1.5{\footnote{The {\sc [O
        II]$\lambda3729/\lambda3727$} intensity ratio in the range
    3.4-1.5 is predicted in photoionization models for the electron
    density in the range ${\it n_e = \rm 10^1-10^5 cm^{-3}}$ for the
    kinetic temperature $T = 10,000 \rm K$
    \citep{Osterbrock2006agna.book.....O}.}}, allowing for the typical
range in the electron density of the gas under photoionization
equilibrium. Gaussian fitting is mainly used to verify any possible
dependences of intensity ratio of \oii\ doublet and its FWHM with
\mgii absorption line properties. However, we simply integrate the
stacked spectrum over the central 12 pixles (i.e., $\sim$800~\kms) for
measuring the \oii\ line luminosity (or \sloii) and the respective
error is computed by propagating the flux uncertainty in each pixel.

We have also generated geometric mean composite spectra to study the
effect of \rrr\ parameter on the average reddening induced by \mgii
absorbers
\citep[][]{York2006MNRAS.367..945Y,Khare2012MNRAS.419.1028K}. For
this, we have shifted each spectrum to the absorber rest frame and
computed their geometric mean. Here, we do not use the normalized or
continuum subtracted spectrum to preserve the average continuum shape
which is important for determining the characteristic extinction law
\citep[see also,][]{York2006MNRAS.367..945Y}. To compute the relative
extinction we have generated the stacked spectra for a control sample
of quasars, within $\Delta z = \pm 0.05$ of \zem and $\Delta r_{\rm
  mag} = \pm 0.5$ of $r_{\rm mag}$, without absorbers in their
spectra. We will discuss the results of this analysis in
Section~\ref{sub:red}.

\section{Results}

We generate several composite spectra in various \ew\ and redshift
bins. Since, we are interested in measuring the average \oii\ nebular
emission line luminosity (or surface brightness) associated with the
\mgii absorbers, the direct detection of \oii\ nebular emission
reported in Paper 1 are also included in most of our analysis. We also
present the results when these systems are excluded from the analysis.

\subsection{Average \oii\ emission and fibre effects}
\label{sub:loiivsew}

 To quantify the fibre loss effect we have constructed composite
 spectra by dividing our sample into five \ew\ bins of $1-1.5,
 1.5-2.0,2.0-2.5, 2.5-3.0$ and $\ge 3$~\AA\ for SDSS-DR7 and DR12. The
 stacked profiles for various \ew\ bins and the number of systems used
 to construct the stacked spectra are shown in Fig.~\ref{fig:profile}.
 It is apparent that for a given equivalent width bin the \oii\ line
 luminosity found for SDSS-DR7 is higher than that of SDSS-DR12. In
 the \emph{left panel} of Fig.~\ref{fig:loiivsew}, we show the median
 \oii\ line luminosity as a function of \ew\ for the \mgii systems in
 SDSS-DR7 (\emph{circles}) and DR12 (\emph{triangles}). We find a
 clear increasing trend of \loii\ with increasing \ew. This is
 consistent with the trend found by
 \citet{Noterdaeme2010MNRAS.403..906N} in a stacking analysis of \mgii
 systems they have found from SDSS-DR7 quasars. We note that median
 \loii\ we detect in SDSS-DR7 sample is similar to that obtained by
 \citet[][see also their table 5]{Noterdaeme2010MNRAS.403..906N}.

In \emph{right panel} of Fig.~\ref{fig:loiivsew}, we show \sloii\ as a
function of \ew. We confirm the strong correlation between \ew\ and
\sloii\ in both the SDSS-DR7 (\emph{circles}) and DR12
(\emph{triangles}) datasets. The \sloii\ measurement for SDSS-DR7 and
DR12 datasets for each \ew\ bin are listed in column 4 and 7 of
Table~\ref{tab:sigloii}, respectively. We first compare our results
based on SDSS-DR7 stacked spectra with that of
\citet{Menard2011MNRAS.417..801M}, based on SDSS-DR4, as spectra were
obtained using 3 arcsec fibre in both cases. Note that our redshift
range of $0.55 \le z \le 1.3 $ is slightly different from that of
\citet{Menard2011MNRAS.417..801M} of $0.36 \le z \le 1.3$. As
suggested by \citet{Menard2011MNRAS.417..801M} we model this
relationship using a powerlaw of the form $\langle$\sloii$\rangle$ =
$A W_0^{\alpha}$ and compute the best fit parameters of $\rm \alpha =
1.69 \pm 0.11$ and $A=$ $ \rm (2.55 \pm 0.27) \times 10^{37} \rm
erg\ s^{-1}\ kpc^{-2}$ for SDSS-DR7. While comparing our best fit
parameters with \citet[][]{Menard2011MNRAS.417..801M}, i.e., $\rm
\alpha = 1.75 \pm 0.11$ and $A=$ $ \rm (1.48 \pm 0.18) \times 10^{37}
\rm erg\ s^{-1}\ kpc^{-2}$, the $\alpha$ is found to be same whereas
normalization factor is found to be different (i.e. $\sim$ 1.7 times
higher) at a significance level of $\sim 3.3\sigma$. We note that this
difference in the normalization factor is mainly due to difference in
the way quasar continuum is modelled.
\citet[][]{Menard2011MNRAS.417..801M} have modelled the continuum with
an iterative running median of sizes ranging from 500 to 15 pixels.
This basically smooths all small scale fluctuations. In fact, we get
the similar values of \sloii\ if we use the continuum fitting
procedure adopted by \citet[][]{Menard2011MNRAS.417..801M} [see
  Fig.~\ref{fig:loiivsew_menard} and related discussions in the
  Appendix]. However, for rest of the paper we will present result
from data using our continuum fitting procedure.

\par

\begin{table*}
 \centering
 \begin{minipage}{150mm}
 {\small
 \caption{\oii\ luminosity surface density traced by \mgii absorbers
   as a function of equivalent width (\ew).}
 \label{tab:sigloii}
\begin{tabular}{@{} c ccccc  c @{}}
 \hline  \hline 
 \multicolumn{1}{c}{\ew}   &\multicolumn{3}{c}{\sloii\ for SDSS-DR7}  &\multicolumn{3}{c}{\sloii\ for SDSS-DR12}    \\
 \multicolumn{1}{c}{interval}   &\multicolumn{1}{c}{$\langle$\ew $\rangle$}  &\multicolumn{1}{c}{$\langle$ $z$ $\rangle$} &\multicolumn{1}{c}{\sloii} &\multicolumn{1}{c}{$\langle$\ew $\rangle$}  &\multicolumn{1}{c}{$\langle$ $z$ $\rangle$} &\multicolumn{1}{c}{\sloii} \\
 \multicolumn{1}{c}{ (\AA)  }   &\multicolumn{1}{c}{        (\AA)          }  &\multicolumn{1}{c}{} &\multicolumn{1}{c}{($\rm  \times 10^{38} erg\ s^{-1}\ kcp^{-2}$)} &\multicolumn{1}{c}{(\AA)}  &\multicolumn{1}{c}{} &\multicolumn{1}{c}{($\rm \times 10^{38} erg\ s^{-1}\ kcp^{-2}$)} \\
\hline

1.0~\AA$ \le $ \ew $ < 1.5$~\AA\ &  1.23    &  0.93      &   $0.28  \pm 0.04 $  & 1.23 &  0.89   &  $0.37 \pm0.04 $ \\
1.5~\AA$ \le $ \ew $ < 2.0$~\AA\ &  1.71    &  0.94      &   $0.68  \pm 0.06 $  & 1.73 &  0.90   &  $0.49 \pm0.05 $ \\
2.0~\AA$ \le $ \ew $ < 2.5$~\AA\ &  2.22    &  0.95      &   $1.07  \pm 0.09 $  & 2.23 &  0.91   &  $0.99 \pm0.08 $ \\
3.0~\AA$ \le $ \ew $ < 3.5$~\AA\ &  2.73    &  0.95      &   $1.71  \pm 0.13 $  & 2.72 &  0.92   &  $1.36 \pm0.11 $ \\
3.5~\AA$ \le $ \ew $ < 6.0$~\AA\ &  3.62    &  0.97      &   $1.99  \pm 0.15 $  & 3.66 &  0.91   &  $1.33 \pm0.12 $ \\
 \hline                                                                                 
 \end{tabular} 
 }             
 \end{minipage}
 \end{table*}

It is clear from the figure that \sloii\ measured from the SDSS-DR7
data are higher than those from SDSS-DR12 data (apart from the first
equivalent width bin). The difference in 3 out of the 5 equivalent
width bin considered here is more than 2$\sigma$ (see column 4 and 7
of Table~\ref{tab:sigloii}). The best fitted relationship between
\sloii\ and \ew\ for the SDSS-DR12 data are shown in \emph{dot-dashed}
line (i.e., $\langle$\sloii$\rangle$ = $A W_0^{\alpha}$ with $\rm
\alpha = 1.28 \pm 0.11$ and $A=$ $ \rm (2.95 \pm 0.29) \times 10^{37}
\rm erg\ s^{-1}\ kpc^{-2}$) in the figure. While this fit is not as
good as that for DR7 data, what is evident is that the fit to DR12
data is consistently lower than that for the DR7 data in all but one
equivalent width bin. This clearly confirms that the observed
\oii\ surface brightness is affected by fibre size effects even after
normalizing the flux with the projected fibre size. In particular our
results are consistent with the fact that (1) there is a large scatter
in the \ew\ vs impact parameter relationship even at high equivalent
widths (see figure 1 of
\citealt{Nielsen2013ApJ...776..114N,Lan2014ApJ...795...31L,Huang2016MNRAS.455.1713H})
and (2) gradients are known to be present in the surface brightness
profiles of galaxies which results in lower measured
\oii\ luminosities when smaller fibres probe predominately the outer
parts of galaxies.

\begin{figure}
  \centering
 \epsfig{figure=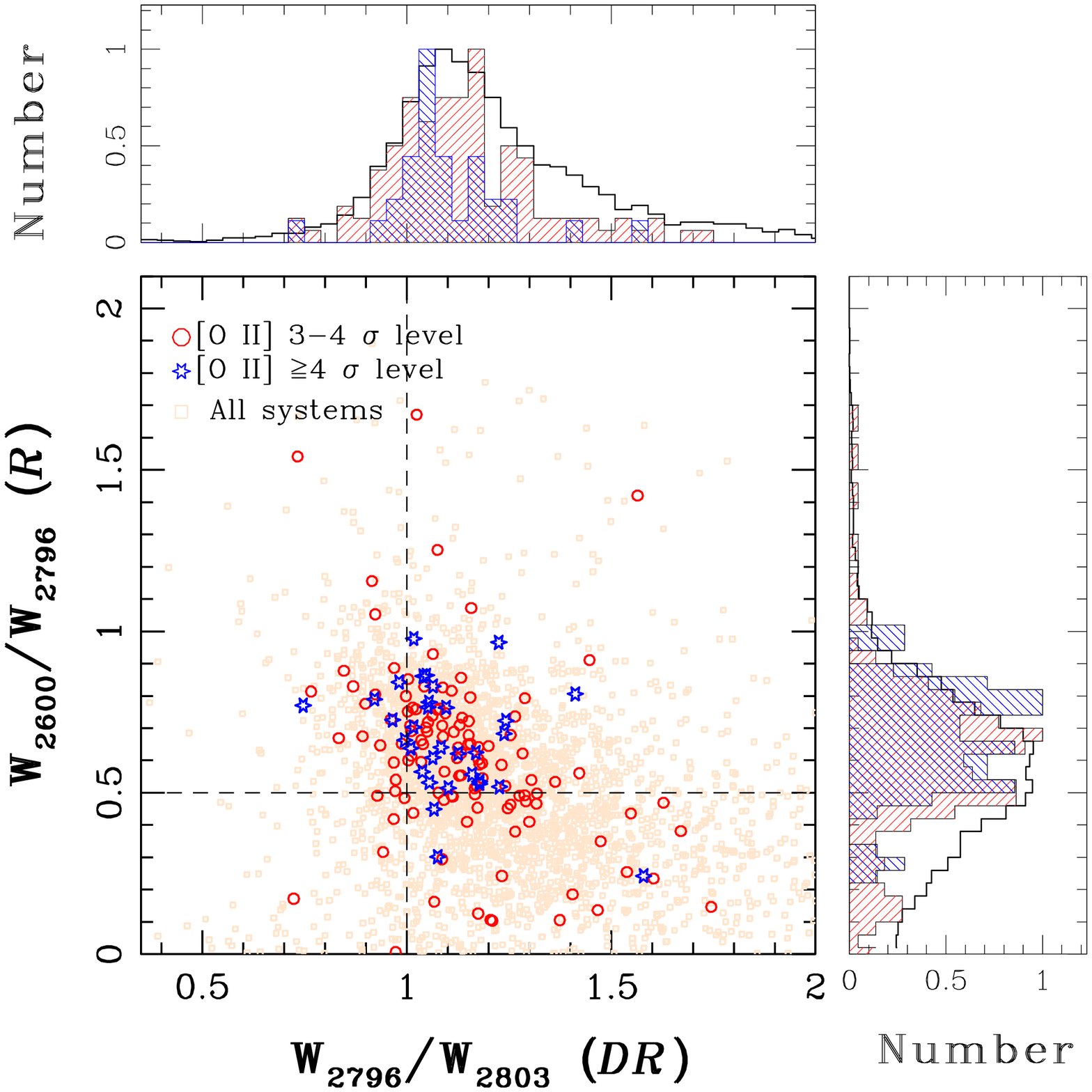,height=8.5cm,width=8.5cm,angle=0}
  \caption{The \mgii doublet ratio versus \rrr\ parameter of \mgii
    absorbers with nebular emission line detected at $\ge 4\sigma$
    (\emph{stars} from Paper 1), $3-4\sigma$ (\emph{circles}) and
    without (i.e., $< 3\sigma$) nebular emission (\emph{squares}) line
    detection for \ew\ $> 1$~\AA\ in SDSS-DR7. The upper and
    right-hand panels show, respectively, the $DR$ and
    \rrr\ distributions for the \mgii systems with nebular emission
    line detected at $\ge 4\sigma$ (shaded with blue slanted lines at
    $-45^{\circ}$), $3-4\sigma$ (shaded with red slanted lines at
    $45^{\circ}$) and without (i.e., $< 3\sigma$) nebular emission
    (unfilled histogram). }
\label{fig:scatter}
 \end{figure}

In the lower panel of Fig.~\ref{fig:sigma_delv} we also show the
dependence of velocity width (deconvolved for the instrumental
broadening) of \oii\ line ($\sigma_{\rm[O II]}$) on the \ew\ of
\mgii\ systems. A clear increasing trend of $\sigma_{\rm[O II]}$ with
\ew\ is apparent from the figure. The $\sigma_{\rm[O II]}$ increases
from $\sim$ 90~\kms\ to $\sim$ 200~\kms\ when the average
\ew\ increases from 1~\AA\ to 3~\AA\ in both SDSS-DR7 and DR12. The
\oii\ line profiles for each \ew\ bin are shown in the upper panel of
Fig.~\ref{fig:sigma_delv}. One can see that emission line peaks for
both the \oiiab\ doublet components are clearly visible in the stacked
profile of lower \ew\ bins, i.e., $< 2.5$~\AA. However, \oiiab\ line
is blended for the strong \mgii systems with \ew\ $> 2.5$~\AA.
Typically $\sigma$ of any emission line from a galaxy can be a good
probe of the underlying mass. However, in the stacked spectrum it can
also be a reflection of spread in the difference between emission and
absorption redshifts. Therefore, a correlation between \ew\ and
$\sigma_{\rm[O II]}$ could either reflects velocity offset ($\Delta
v$) between emission and absorption increasing with \ew\ and/or high
\ew\ systems originating from massive halos. However, in Paper 1 when
we considered the direct \oii\ detections we do not find any
correlation between \ew\ vs $\sigma_{\rm[O II]}$ as well as \ew\ vs
$\Delta v$ (relative velocity between absorption and emission
redshift). It is also clear from Fig.~\ref{fig:sigma_delv}, that
$\sigma_{\rm[O II]}$ measured for a given \ew\ bin matches very well
between SDSS-DR7 and SDSS-DR12. In addition, we do not find any
dependence of \oiiab\ doublet ratio with \ew.

 It is clear from the above discussions that \loii\ and
 \sloii\ measured in the SDSS-DR12 composite are under estimated.
 Therefore, to minimize the fibre size effects, in most discussions
 that follows we concentrate on results based on DR7.

\subsection{Dependence of \oii\ emission on \rratio}
\label{sec:rpara}

   In Paper 1, we have shown that for a given luminosity threshold
   (i.e. \loii), direct \oii\ nebular line detection fraction
   increases with increasing \ew. We have considered \oii\ detections
   with more than 4$\sigma$ significance level in that study. Here,
   before doing the stacking analysis, we also identify systems with
   tentative emission feature (hereinafter, candidate \oii\ emitters)
   at the expected position of \oii\ at $ 3 < \sigma < 4$
   level{\footnote{Please refer to paper 1 for how we compute the
       significance level for the \oii\ feature.}}. Note that by
   lowering the significance level to confirm a detection, we might
   have enhanced the number of false positives. In
   Fig.~\ref{fig:scatter}, we compare the distribution of these \mgii
   systems with (\emph{circles}) and without (\emph{squares}) emission
   feature in the {\rm {\ensuremath{\mathcal{R}}}} vs $DR$ plane
   (systems discussed in Paper 1 marked with star symbol).
   Interestingly, like the trend shown by firm detections, most of the
   \mgii systems with consistent features (at $3-4 \sigma$ level) at
   the location of \oii\ nebular line also have {\rm
     {\ensuremath{\mathcal{R}}}} $\ge 0.5$ and $DR$ $\sim$ 1 (see
   also, figure 8 in Paper 1). In Fig.~\ref{fig:scatter}, we also plot
   the histogram of {\rm {\ensuremath{\mathcal{R}}}} and $DR$
   distribution in the right and upper panel respectively. The
   sub-samples of ``candidate \oii\ emitters'' and systems without
   nebular emission are found to be drawn from different distribution
   of {\rm {\ensuremath{\mathcal{R}}}} with $KS-test$ null probability
   of $p_{null} = 0.004$. It again indicates that the luminosity of
   \oii\ emission in the stacked spectrum will depend on {\rm
     {\ensuremath{\mathcal{R}}}} parameter.

  \begin{figure*}
    \centering 
  \epsfig{figure=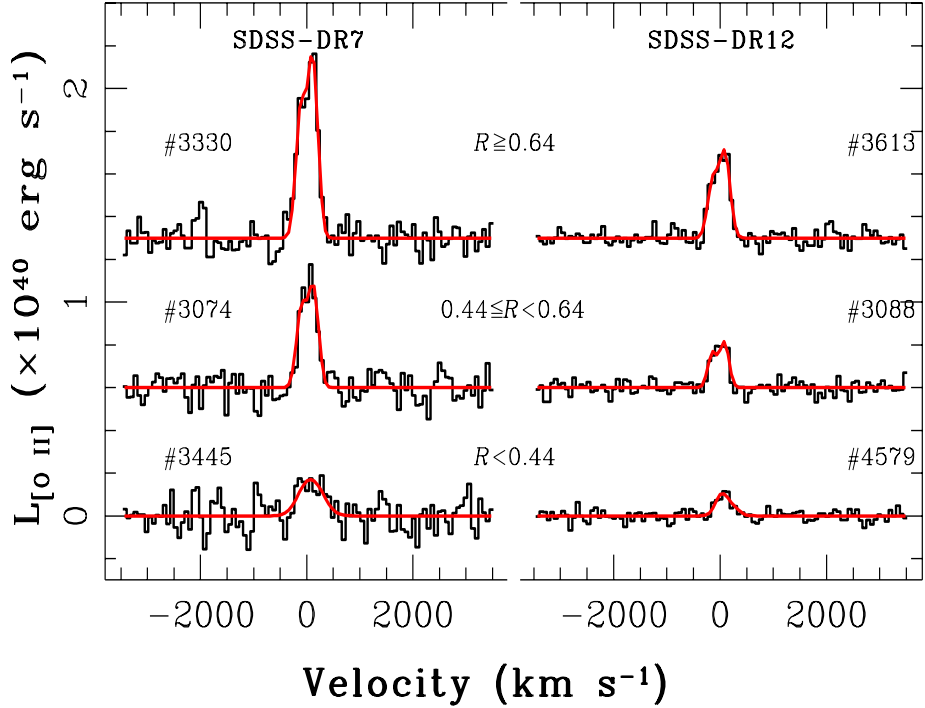,height=10cm,width=12.0cm,bbllx=37bp,bblly=151bp,bburx=545bp,bbury=517bp,clip=true}
   \caption{\oii\ luminosity (\loii) in the stacked
     spectra obtained for three bins of \rratio\ for strong (\ew $>
     1$~\AA) \mgii absorbers detected in SDSS-DR7 spectra. \emph{Right
       panel:} The same for the \mgii absorber detected in SDSS-DR12
     spectra. Total number of \mgii\ systems involved in each bin is
     also given in the figure.}
 \label{fig:stackfeiiall}
  \end{figure*}

\begin{table*}
 \centering
 \begin{minipage}{110mm}
 {\scriptsize
 \caption{The \loii\ for the subset based on \rrr\ parameter.}
 \label{tab:stat}
\begin{tabular}{@{} c ccccc  c @{}}
 \hline  \hline 
 \multicolumn{1}{c}{Criteria}   &\multicolumn{5}{c}{SDSS-DR7}    \\
                                &\multicolumn{1}{c}{No}  &\multicolumn{1}{c}{$\langle$\ew $\rangle$} &\multicolumn{1}{c}{$\langle z \rangle$} &\multicolumn{1}{c}{\loii}&\multicolumn{1}{c}{\loii$^P$\textcolor{blue}{$^a$}} \\
                                &\multicolumn{1}{c}{  }  &\multicolumn{1}{c}{(\AA)} &\multicolumn{1}{c}{                   } &\multicolumn{1}{c}{($\rm \times 10^{40}~erg~s^{-1}$)} &\multicolumn{1}{c}{($\rm \times 10^{40}~erg~s^{-1}$)}\\ 
\hline

1~\AA$ \le $ \ew $ < 2$~\AA\ &      &        &         &                    &   \\

 \rrr $\le  0.44$          & 2871 &   1.36 &    0.95 &    $0.99 \pm   0.26$ & $1.66  $\\
 $0.44 < $ \rrr $\le 0.64$ & 2169 &   1.42 &    0.93 &    $1.48 \pm   0.28$ & $1.79  $\\
 \rrr $> 0.64$             & 2291 &   1.43 &    0.92 &    $2.76 \pm   0.26$ & $1.82  $\\

                           &      &        &         &                      &     \\
 \hline                                                                           
2~\AA$ \le $ \ew $ < 3$~\AA\ &      &        &         &                    &      \\

  \rrr $\le  0.44$          &450 &   2.35 &   0.95 &  $1.52  \pm    0.69$  & $4.40  $ \\
  $0.44 < $ \rrr $\le 0.64$ &675 &   2.41 &   0.95 &  $4.29  \pm    0.59$  & $4.60  $ \\
  \rrr $> 0.64$             &801 &   2.41 &   0.94 &  $7.94  \pm    0.48$  & $4.60  $ \\
 \hline                                                                                 
 \end{tabular} 
 }             
\\
\textcolor{blue}{$^a$}{ expected \oii\ luminosity measured  from the \ew\ vs \loii\ \\ correlation for the average \ew\ per bin.}
 \end{minipage}
 \end{table*}

\begin{figure*}
  \centering
  \epsfig{figure=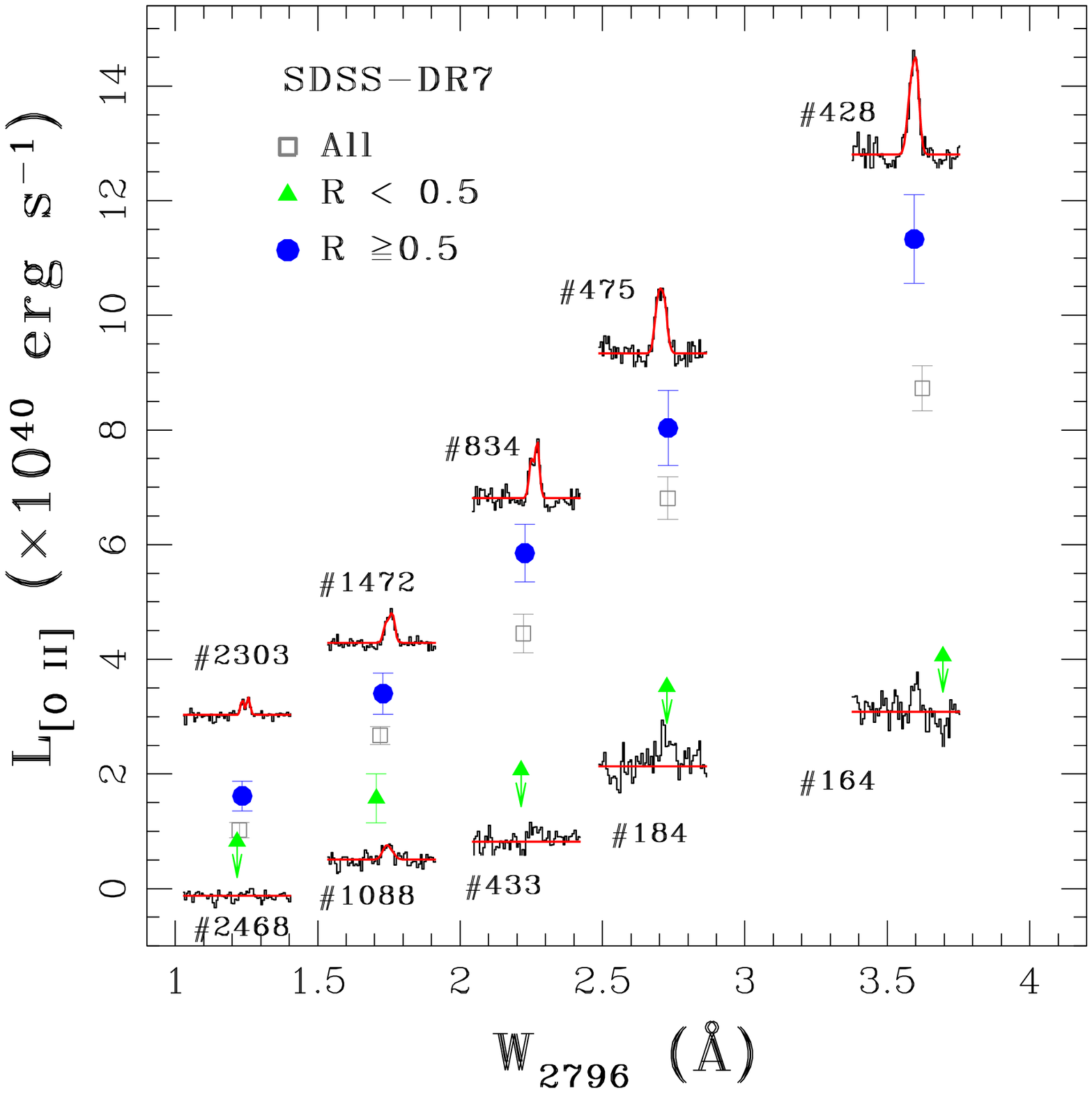,height=8.3cm,width=8.3cm,angle=0}
  \epsfig{figure=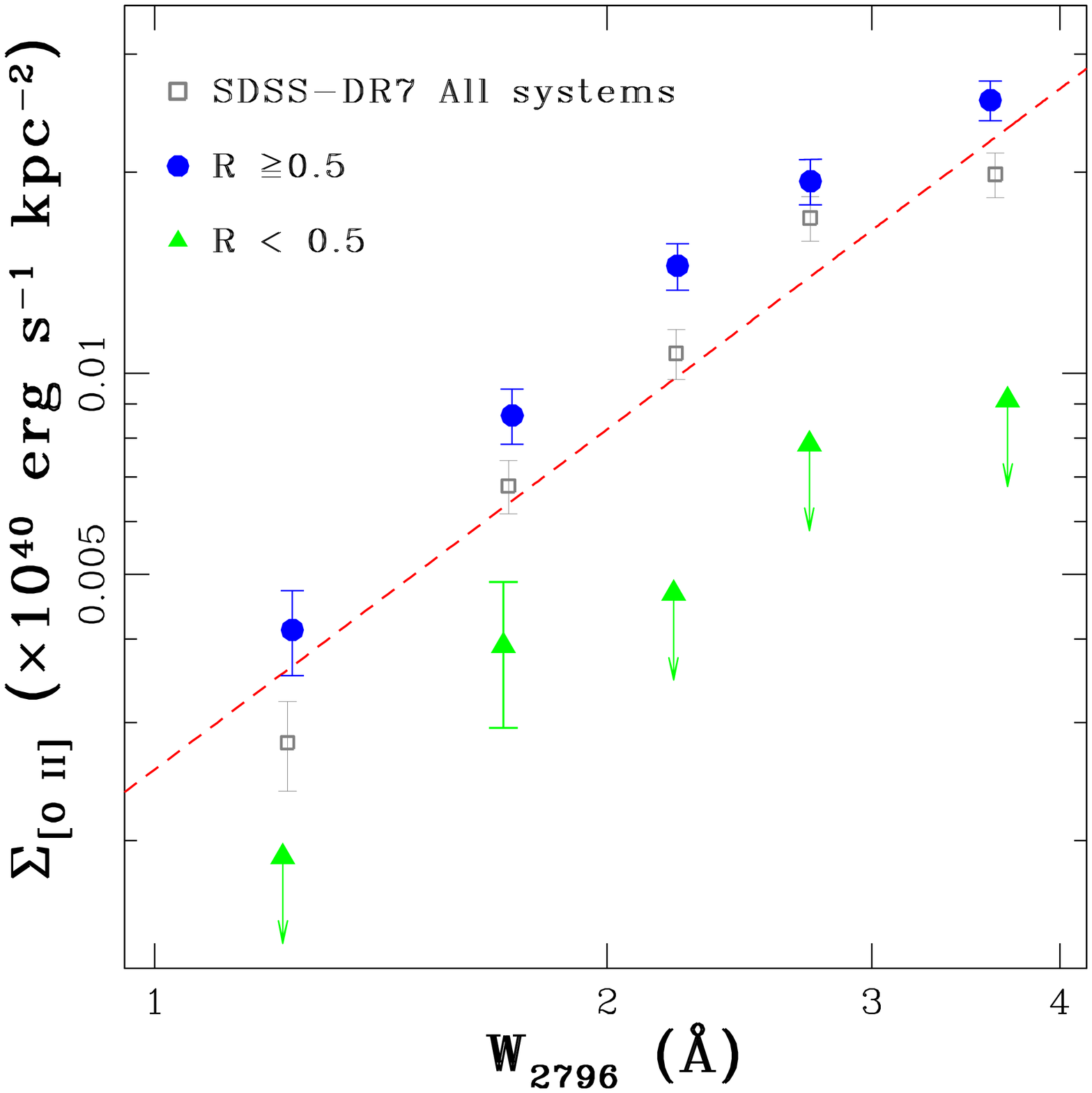,height=8cm,width=8cm,angle=0}
  \caption{\emph{Left panel:} The \oii\ luminosity of \mgii systems
    detected in SDSS-DR7 with $R \ge 0.5$ (\emph{circles}) and $R < 0.5$
    (\emph{triangles}) for various \ew\ bins. The number of systems
    used for the stack and the stacked profiles are shown. The
    \loii\ for all the systems for different \ew\ bins are shown as
    \emph{squares}. \emph{Right panel:} The \oii\ luminosity surface
    density (\sloii) as a function of \ew. The symbols are as in left
    panel. (\emph{circles}). }
\label{fig:loiivsew_r}
 \end{figure*}

\begin{figure}
  \centering
  \epsfig{figure=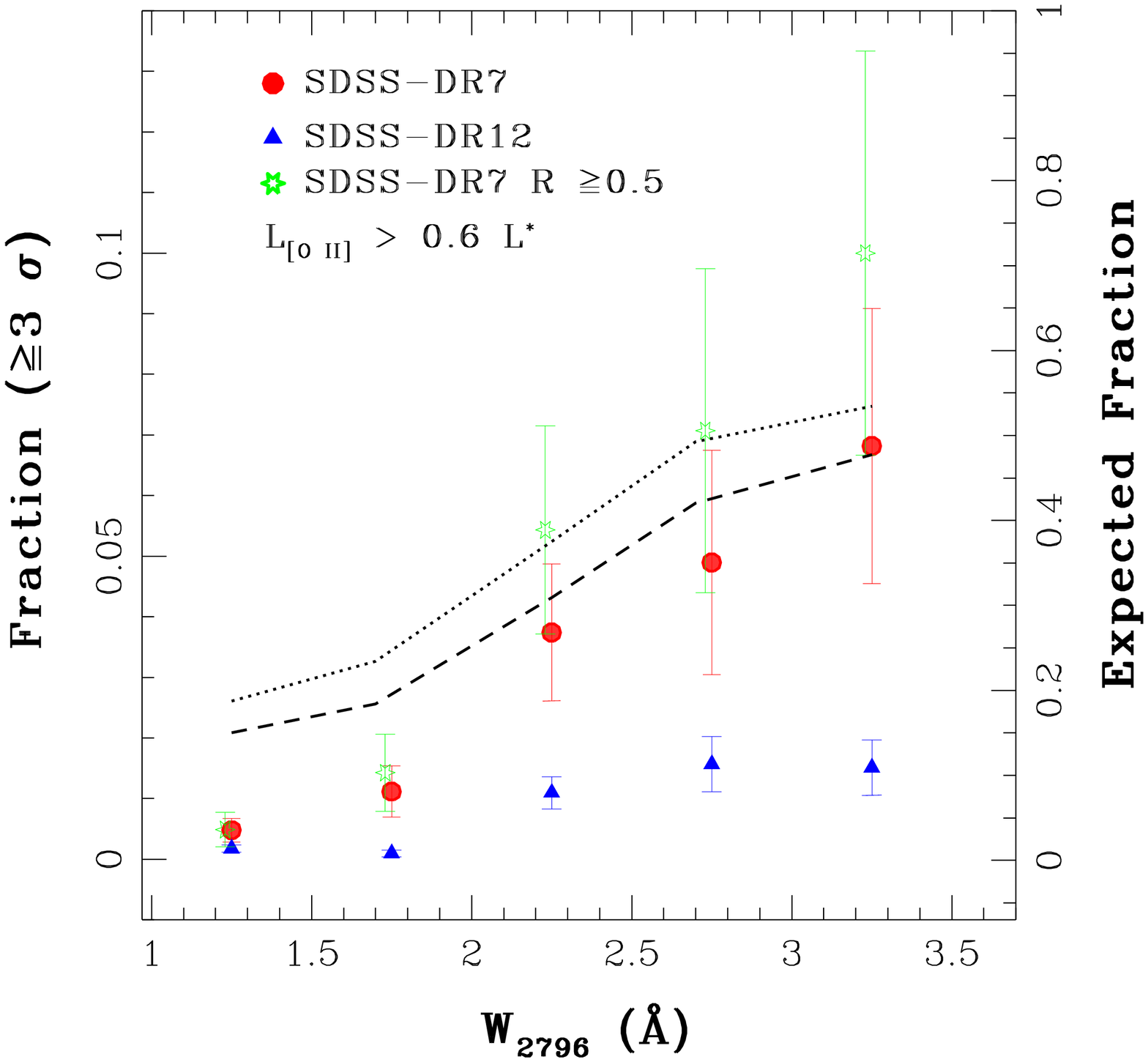,height=7.8cm,width=7.8cm,angle=0}
  \caption{The fraction of \mgii systems with \oii\ nebular emission
    detected at $\ge 3\sigma$ in SDSS-DR7 (\emph{circles}) and DR12
    (\emph{triangles}). The \emph{dotted} and \emph{dashed} lines show
    the expected fraction of \mgii\ systems detected within a
    projected fibre radius of 10 and 7 kpc at a redshift of 0.6, from
    \ew\ vs $\rho$ distribution of confirmed \mgii galaxies by
    \citet{Nielsen2013ApJ...776..114N}, given in the right side
    ordinates. }
  \label{fig:loiivsfrac}
 \end{figure}

 To explore the dependence of average \loii\ and \sloii\ emission on
 {\rm {\ensuremath{\mathcal{R}}}} parameter we generate the composite
 spectra based on {\rm {\ensuremath{\mathcal{R}}}} by dividing our
 sample into three bins of {\rm {\ensuremath{\mathcal{R}}}} $\le
 0.44$, $0.44 < $ {\rm {\ensuremath{\mathcal{R}}}} $\le 0.64$ and {\rm
   {\ensuremath{\mathcal{R}}}} $> 0.64$, having almost equal number of
 systems in each bin. In Fig.~\ref{fig:stackfeiiall}, we show the
 median \oii\ emission line profile in sub-samples based on
 \rrr\ parameter for SDSS-DR7. For comparison, we also show the
 stacked profiles from SDSS-DR12. It is apparent that the strength of
 \oii\ emission is higher for the systems with higher \rrr\ parameter.
 This trend is consistent with the fact that when \oii\ nebular
 emission is detected (with $\ge 3 \sigma$ level of significance) in
 individual cases one finds \rrr\ values to be higher (see
 Fig.~\ref{fig:scatter}).

In order to decipher the effect of \rrr\ (without being affected by
\ew\ vs \loii\ correlations) on the strength of \oii\ luminosity (or
\sloii) we restrict ourselves to two narrow \ew\ range of $\rm 1~\AA
\le $\ew$ < 2$~\AA\ and $\rm 2~\AA \le $\ew$ < 3$~\AA, having
sufficient number of \mgii systems with \rrr $< 0.5$ (see
Fig~\ref{fig:fraction}). Further, we divide each of them in to three
bins of \rrr\ $\le 0.44$, $0.44 < $ \rrr\ $\le 0.64$ and \rrr\ $>
0.64$. The details of these sub-samples are given in
Table~\ref{tab:stat}. The number of systems in each sub-sample,
average \ew\ and \zabs\ are given in column 2, 3 and 4 of this table,
respectively. It is clear that these quantities do not differ by a
wide margin between the sub-samples. However, we clearly see an
increasing trend in \loii\ with \rrr.

A two-sided Kolmogorov-Smirnov test (KS − test) shows that the
\ew\ distribution of two sub-samples with higher {\rm
  {\ensuremath{\mathcal{R}}}} values, i.e., $0.44 < $ {\rm
  {\ensuremath{\mathcal{R}}}} $\le 0.64$ and {\rm
  {\ensuremath{\mathcal{R}}}} $> 0.64$, are drawn from a parent
distribution with a null probability of being drawn from same parent
distribution of $P_{KS} =$ 0.79 and $P_{KS} =$ 0.87 for both
\ew\ bins. The systems with {\rm {\ensuremath{\mathcal{R}}}} $\le
0.44$ show slightly lower \ew. The average \ew\ is lower by 4\% in
this case (i.e., {\rm {\ensuremath{\mathcal{R}}}} $\le 0.44$) compare
to other two cases. We find that the probability for sub-samples with
{\rm {\ensuremath{\mathcal{R}}}} $\le 0.44$ and {\rm
  {\ensuremath{\mathcal{R}}}} $> 0.64$ are drawn from two different
populations is $P_{KS} = 1.7 \times 10^{-11}$ (respectively, 0.03) for
the systems having $\rm 1~\AA \le $\ew$ < 2$~\AA\ (respectively, $\rm
2~\AA \le $\ew$ < 3$~\AA).

Next, we ask whether difference in \loii\ between \rrr\ $\le 0.44$ and
\rrr\ $> 0.64$ comes from the previously discussed correlation between
\ew\ and \loii. Based on our \loii\ vs \ew\ best fit parameters (i.e.,
$A W_0^{\alpha}$ with $\rm \alpha = 0.97 \pm 0.07$ and $A=$ $ \rm
(1.78 \pm 0.08) \times 10^{40} \rm erg\ s^{-1}$) we compute the
expected \loii\ for the mean \ew\ probed by the above sub-samples and
is given in column 6 of Table~\ref{tab:stat}. It is clear from this
table that purely based on the \ew\ vs \loii\ we do not expect
\loii\ to be very different between different sub-samples.
Interestingly, for a fixed \ew\ range of $\rm 1~\AA \le $\ew$ <
2$~\AA\ a clear difference in \loii\ of factor 2.8 (significant at 3.6
$\sigma$) is apparent for the subset having different \rrr\ parameter
of \rrr\ $\le 0.44$ and \rrr\ $> 0.64$. The difference in \loii\ is
found to be even higher of about factor 5.7 (significant at
4.8$\sigma$) if we consider the systems with $\rm 2~\AA \le $\ew$ <
3$~\AA.

\emph{The discussions presented here confirm that \loii\ obtained in
  the stacked spectra depends strongly on \rrr} \emph {parameter. }

\subsection{Dependence of  \loii\ versus \ew\ based on  \rrr}
 \label{sub:loiivsr}

  In this section we explore the effect of \rrr\ on \loii\ vs
  \ew\ relation using SDSS-DR7 data. For this, we generate composite
  spectra for two subsets with \rrr\ $\ge 0.5$ and \rrr\ $< 0.5$ for
  various \ew\ bins. In the left panel Fig.~\ref{fig:loiivsew_r}, we
  plot the \loii\ vs \ew. We also show the \oii\ profile for each
  \ew\ bin for two ranges in \rrr. It is clear from the figure that
  the \oii\ emission has systematically higher luminosity in the case
  of \rrr\ $\ge 0.5$ in each \ew\ bin which is consistent with our
  findings discussed above. For the subset with \rrr\ $< 0.5$ we note
  that even for the lower \ew\ bins, where there are a good number of
  systems available for the stacking, either the emission is
  significantly weaker or is not detected. \par

In right panel of Fig.~\ref{fig:loiivsew_r}, we show that the median
surface luminosity density for the systems with \rrr\ $\ge 0.5$
(\emph{open circles}) is higher than those of all \mgii systems
(\emph{solid circles}). It is clear from this figure that the
correlation, we as well \citet[][]{Menard2011MNRAS.417..801M}, found
between \loii\ and \ew\ in the whole \mgii sample is mainly dominated
by systems showing strong Fe{\sc~ii}. For systems with \rrr\ $< 0.5$
no clear trend between \ew\ and \loii\ is visible and we could measure
only upper limits in several \ew\ bins.

 Note, contrary to the correlation seen between \ew\ and \loii\ (or
 \sloii) in the stacked spectra individual detections do not follow
 this correlation. In Paper 1, we suggested that the \ew\ vs
 \loii\ correlation may come from increase in \oii\ detection fraction
 with increasing \ew. We explore this point further here.
 
In Fig.~\ref{fig:loiivsfrac}, we show the fraction of \mgii systems
with an emission feature detected at $\rm \ge 3\sigma$ at the expected
location of \oii\ doublet, having \ew\ threshold of $\ge 1$~\AA\ and
\loii\ $\ge 0.6$~\lsoii. A similar plot for \mgii systems with
\oii\ nebular emission detected at $\ge 4 \sigma$ ($\sim$ 198 systems)
are presented in Paper 1. Interestingly, we find a clear increasing
trend between the fraction of \mgii systems with \oii\ nebular
emission and \ew\ even among the tentative detections. A similar trend
in also seen when we consider the fraction of systems with detection
threshold of $\ge 2\sigma$ as well. This can be naturally explained
with the known anti-correlation between \ew\ and $\rho$ where the
galaxies at lower projected distances (i.e., impact parameters)
produce on an average stronger \mgii absorption and higher probability
of the associated \oii\ emission falling inside the fibre. The
fraction further increases if we put an additional constraint of
\rrr\ $\ge 0.5$. The reason being, as discussed in
Fig.~\ref{fig:scatter} most of the direct detections have \rrr\ $\ge
0.5$. \par

  Using the impact parameter distribution of 183 spectroscopically
  confirmed \mgii galaxies from the compilation of
  \citet{Nielsen2013ApJ...776..114N} we compute the detection fraction
  of \mgii\ systems within the impact parameter of $\sim 10$ and $\sim
  7$ kpc for SDSS-DR7 and DR12, i.e., the projected radius of 3 and 2
  arcsec fibres at $z=0.6$. For this, we measure the average $\rho$
  and its standard deviation from the $\rho$ distribution of \mgii
  systems for various \ew\ bins. For each \ew\ bin we compute the
  probability of galaxy to come inside the fibre by randomly
  generating 10,000 values of $\rho$ by assuming a Gaussian
  distribution and considering $1 \sigma$ error over each measurement.
  In Fig.~\ref{fig:loiivsfrac}, we show the expected fraction of \mgii
  systems in SDSS-DR7 (\emph{dotted line}) and DR12 (\emph{dashed
    line}) which roughly follow the observed trend in our systems.
  This once again reiterates the importance of anticorrelation between
  \ew\ vs $\rho$ and the fibre losses in deriving the correlations
  seen in the stacked spectra. However, it is important to note that
  the sample of \citet{Nielsen2013ApJ...776..114N} does not contain
  enough systems at $\rho\le10$ kpc. As discussed before at such
  impact parameters the relationship between \ew\ and $\rho$ may not
  be similar to what we see at high $\rho$ values. Thus it is very
  important to quantify the extent and nature of star forming regions
  associated with strong Mg~{\sc ii} systems at low impact parameter
  through direct observations.

  \par

 Our results also suggest that most of the \mgii absorbers with
 \rrr\ $\ge 0.5$ should have systematically lower impact parameter (or
 higher \loii\ and hence higher star formation rate) compared to those
 with \rrr\ $< 0.5$. We could not check this with the existing data of
 \citet{Nielsen2013ApJ...776..114N} as \feiia measurements are not
 available for most of the systems.

\subsection{ Dependence of \loii\ versus $z$ based on \rrr}
 \label{sub:loiivsz}

The strong correlation between \sloii\ and \ew\ is also found to be
evolving with redshift in the sense that a system with a given
\ew\ seems to be associated with larger \loii\ at high redshift
compared to that at low redshift \citep{Menard2011MNRAS.417..801M}.
Such a redshift evolution of \loii\ is also seen among direct
detections discussed in Paper 1. Here, to explore the dependence of
\loii\ as a function of redshift based on \rrr\ parameter we make two
subsets with \ew\ bins of 1~\AA\ $\le$ \ew $<2$~\AA\ and 2~\AA\ $\le$
\ew $<6$~\AA. We further divide each subset in to three redshift bins
of $0.55 \le z < 0.75$, $0.75 \le z < 0.95$ and $0.95 \le z < 1.3$,
respectively. For these redshifts ranges, the fibre of 3 arcsec
diameter used in SDSS-DR7 projects an angular size of $6.4-7.3$ kpc,
$7.3-7.9$ kpc and $7.9-8.4$ kpc, respectively, in the sky. \par

In Fig.~\ref{fig:loiivsz}, we show the \loii\ versus redshift as a
function of \rrr\ for the subset with 1~\AA\ $\le$ \ew\ $<
2$~\AA\ (\emph{top left panel}) and 2~\AA\ $\le$ \ew\ $<
6$~\AA\ (\emph{top right panel}). It is clear from the figure that the
median \loii\ of \mgii systems is higher at high redshifts. Here also,
we note that the systems with \rrr\ $< 0.5$ show very less emission at
each redshift bin, albeit having similar number of systems as in the
sub-sample of $\rrr$ $\ge 0.5$. The average \loii\ probed in the
stacked spectra corresponds to sub-\lsoii\ [with log \lsoii $\rm
  (erg\ s^{-1}) = 41.60$ at the median $z$ of 0.65] galaxies with
\oii\ luminosity of $\sim 0.01$\lsoii\ and $\sim 0.1$\lsoii, for the
systems with \ew\ ranging from 1~\AA\ $\le$ \ew $<2$~\AA\ and
2~\AA\ $\le$ \ew $<6$~\AA, respectively. The \emph{dashed curve} in
Fig.~\ref{fig:loiivsz} shows the expected luminosity of $0.1$ and
$0.03$ \lsoii\ galaxy as a function of redshift using the redshift
evolution of field galaxies luminosity function by \citet[][see their
  Table 7]{Comparat2016MNRAS.461.1076C}. Note that, these average
luminosities are smaller than the direct detection of \oii\ emission
associated to individual \mgii systems (see figure 9, 10 of Paper 1)}.
  It is clear from Fig.~\ref{fig:loiivsz} that the increase in
  \loii\ associated with \mgii systems roughly follows the luminosity
  evolution of field galaxies and is mostly due to the systems with
  \rrr\ $\ge 0.5$.

 \begin{table}
 \centering
 {\scriptsize
 \caption{The best-fit parameters for the \sloii\ as a function of  redshift, $A (1+z)^{\alpha}$.}
 \label{tab:sample}
 \begin{tabular}{@{} c c c   c @{}}
 \hline  \hline 
 \multicolumn{1}{c}{\ew}    &\multicolumn{2}{c}{SDSS-DR7 with $R \ge 0.5$}  \\
 \multicolumn{1}{c}{(\AA)}  &\multicolumn{1}{c}{A} & \multicolumn{1}{c}{$\alpha$} \\
\hline

 1~\AA\ $\le$ \ew\ $<$ 2~\AA\       &  0.005$\pm$0.003 &    0.01$\pm$0.84 \\
 2~\AA\ $\le$ \ew\ $<$ 6~\AA\       &  0.010$\pm$0.002 &    0.98$\pm$0.39 \\

 \hline                                                                                 
 \end{tabular} 
 }             
 \end{table}

Next, to account for the effect of increasing fibre size with redshift
we also plot \sloii\ as a function of redshift for two equivalent
width bins of 1~\AA\ $\le$ \ew $<2$~\AA\ (\emph{lower left panel}) and
2~\AA\ $\le$ \ew $<6$~\AA\ (\emph{lower right panel}). For
1~\AA\ $\le$ \ew $<2$~\AA\ the \sloii\ seem to be constant with
redshift. However, a mild increase in \sloii\ with redshift is seen
for the subset of 2~\AA\ $\le$ \ew $<6$~\AA. We model \sloii\ vs $z$
with a power-law of the form $A (1+z)^{\alpha}$. The best fit
parameters, i.e., normalization and slope for above two \ew\ bins are
given in column 2 and 3 of Table~\ref{tab:sample}, respectively. We
note that the evolution of \loii\ with redshift is similar (within
$1\sigma$) for both the subsets (see column 3 of
Table~\ref{tab:sample}).

\begin{figure*}
  \centering
  \epsfig{figure=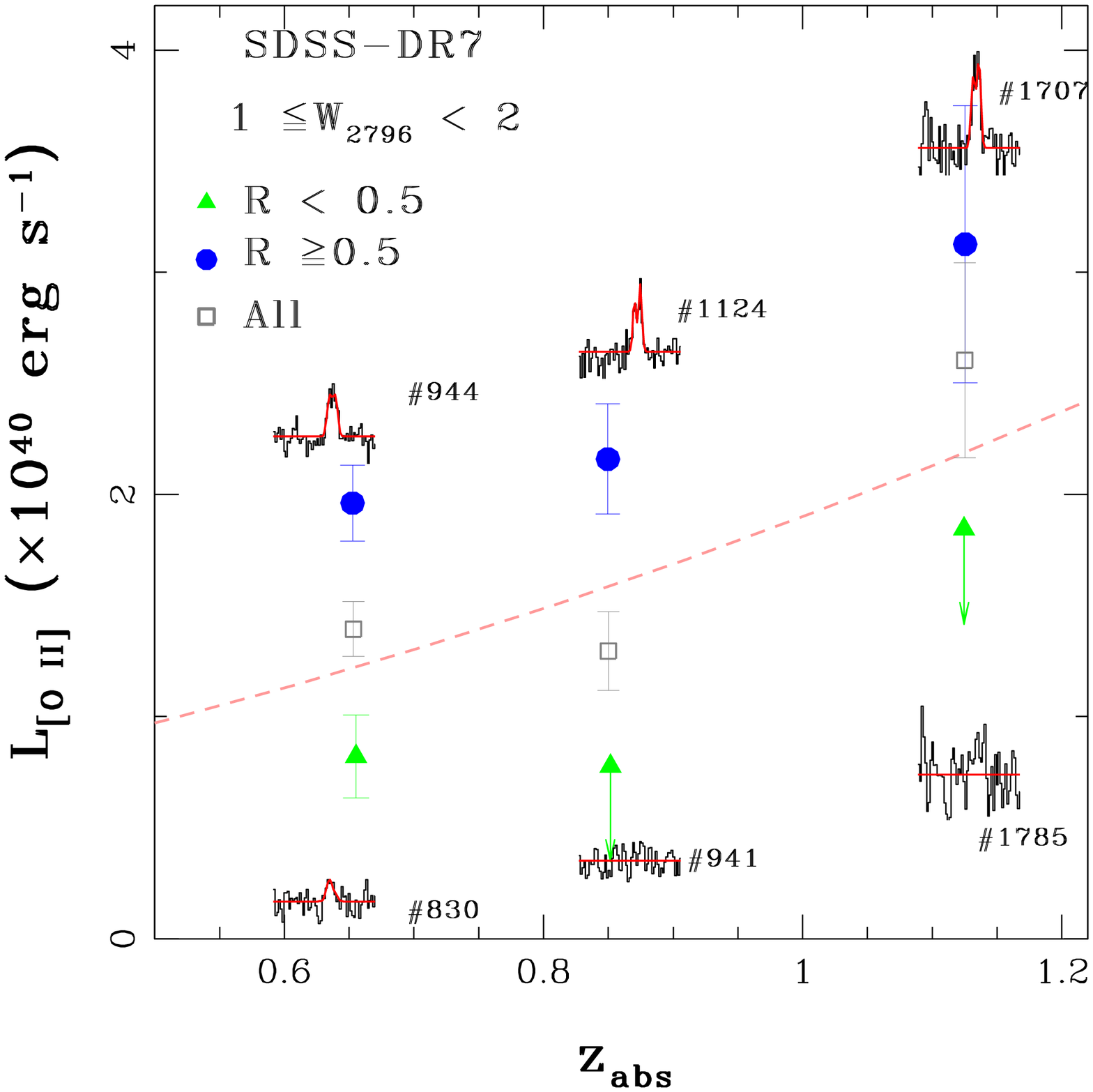,height=7.0cm,width=7.0cm,angle=0}
 \epsfig{figure=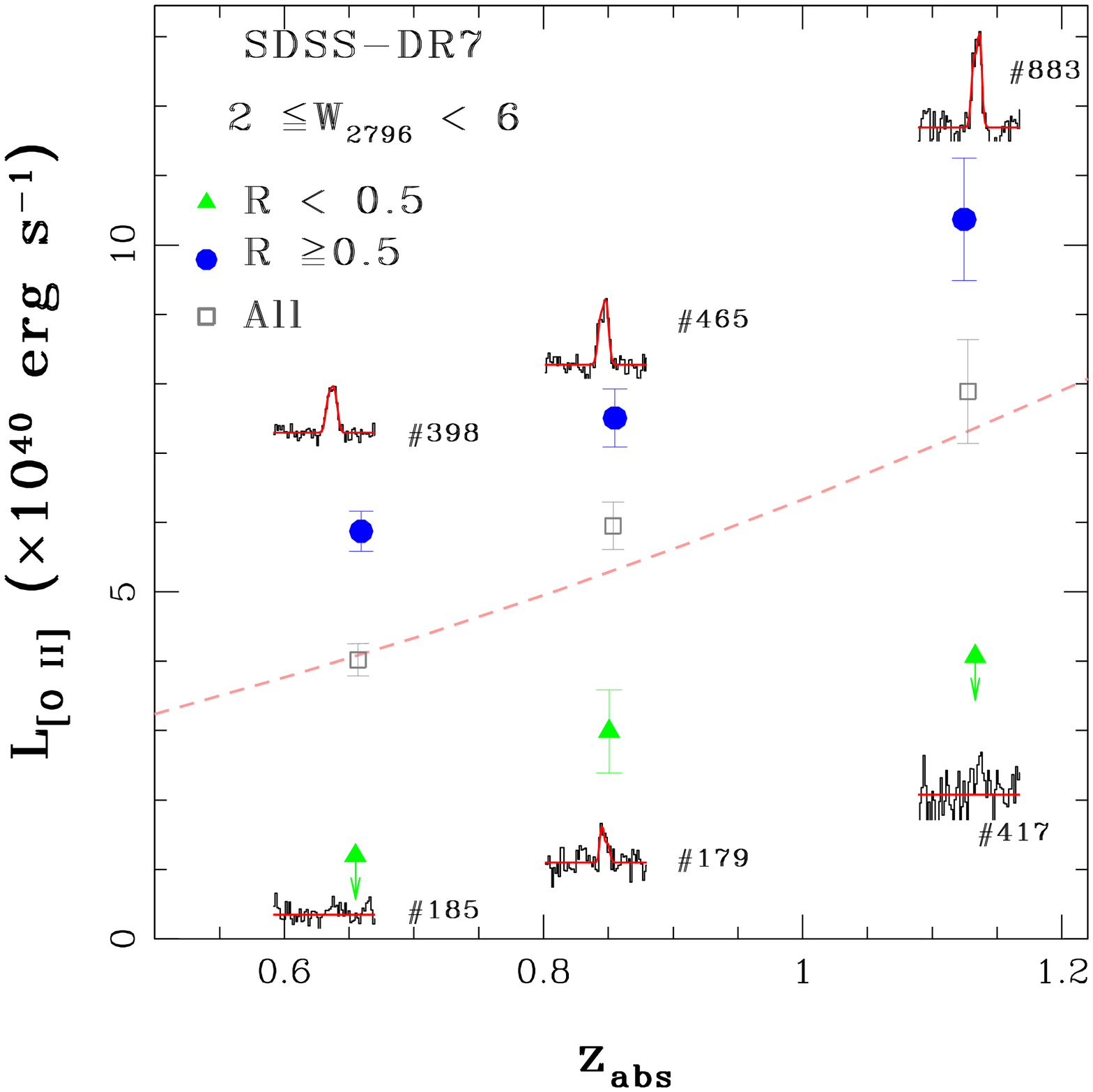,height=7cm,width=7cm,angle=0}
 \epsfig{figure=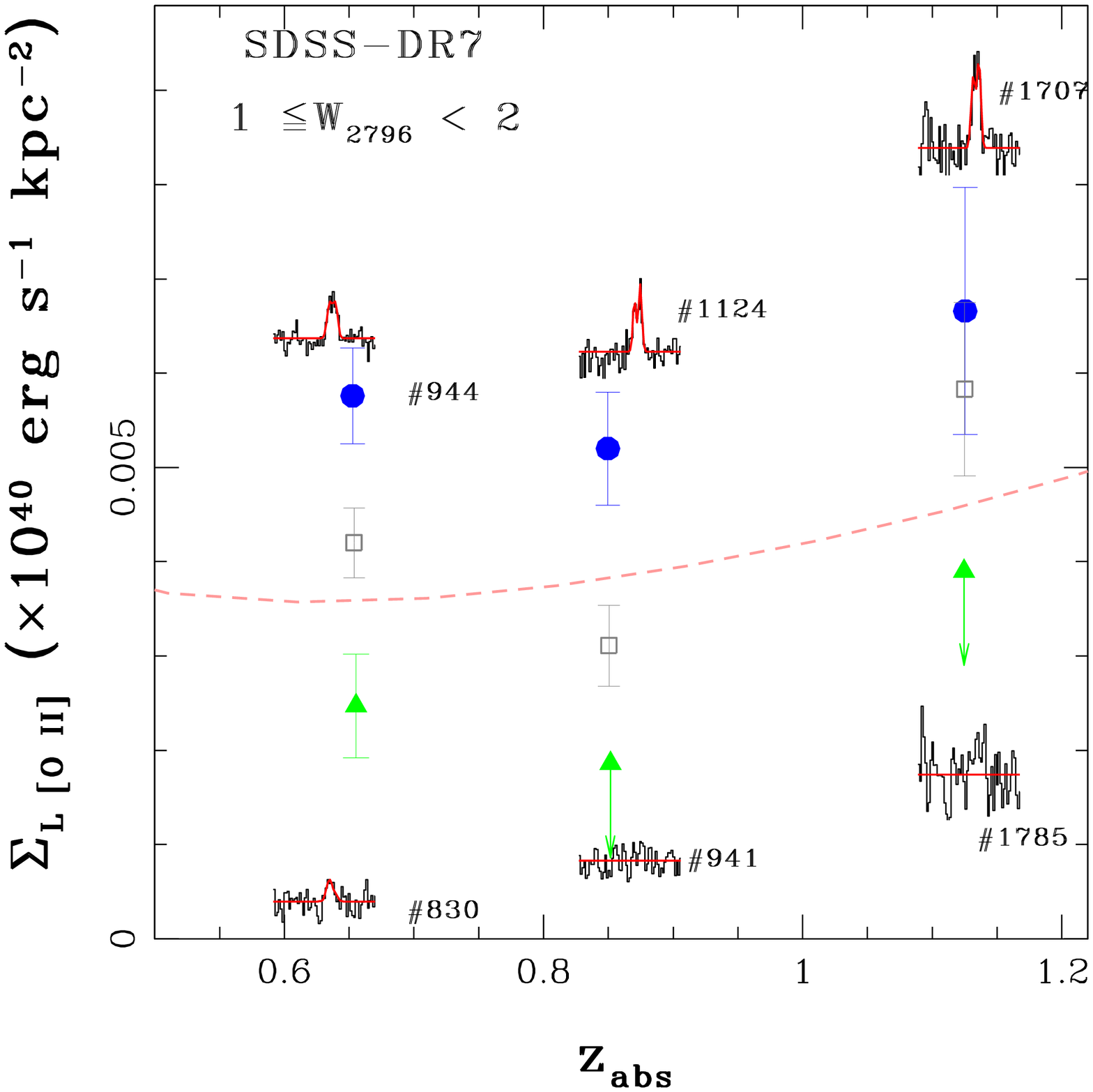,height=7.0cm,width=7.0cm,angle=0}
 \epsfig{figure=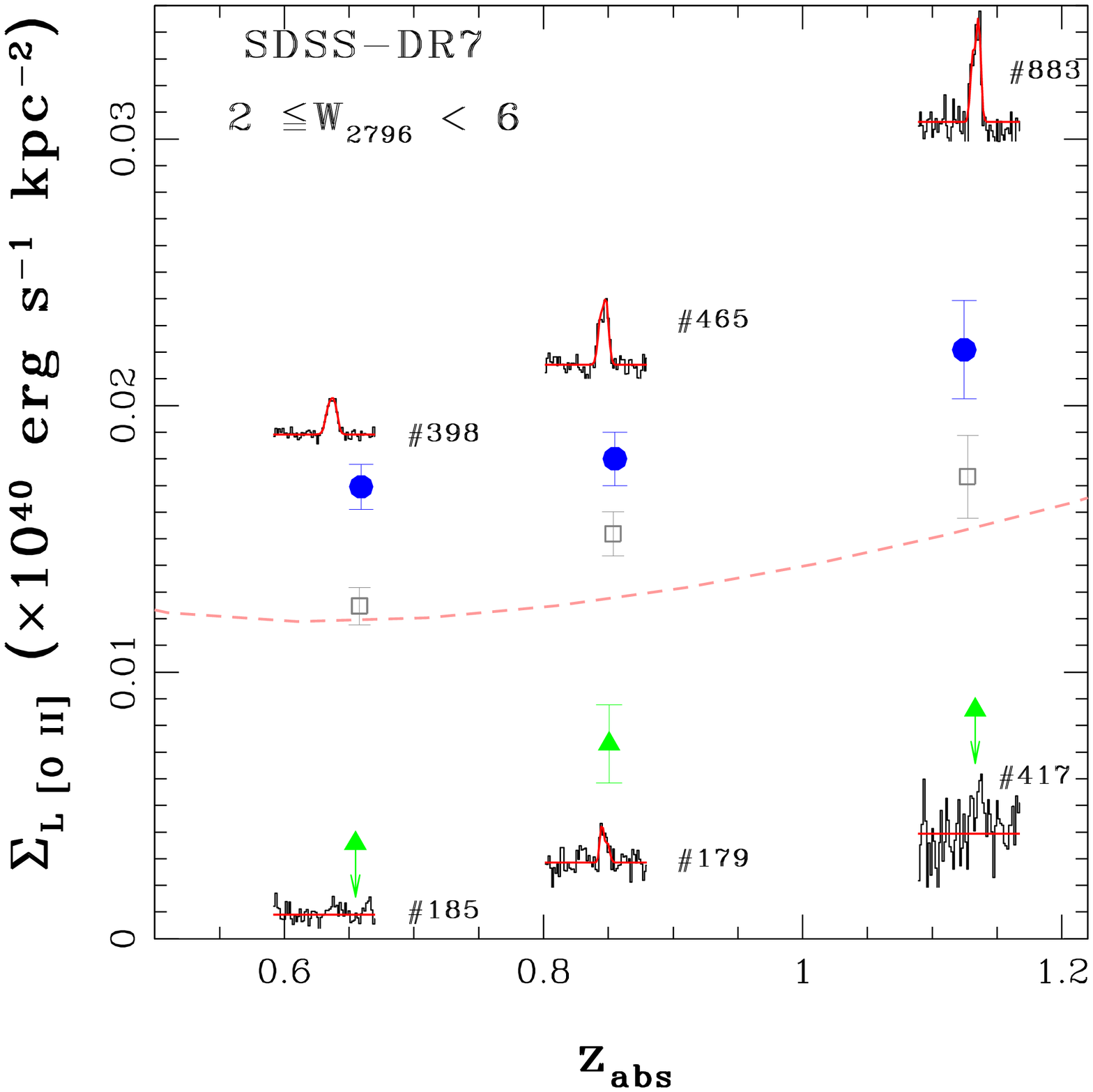,height=7cm,width=7cm,angle=0}

  \caption{\emph{Top left panel:} The \oii\ luminosity as a function
    of \zabs\ for a subset of \mgii systems from SDSS-DR7 with
    1~\AA\ $\le$ \ew $<2$~\AA\ and $R \ge 0.5$ (\emph{circles}) and $R
    < 0.5$ (\emph{triangles}), respectively. The \loii\ for all the
    systems is shown as \emph{squares}. The number of systems used for
    the stack and the stacked profiles are shown. The \emph{dashed
      line} show the luminosity evolution of 0.03\ls (\emph{left
      panel}) and 0.1\ls (\emph{right panel}) galaxy as a function of
    redshift. \emph{Top right panel:} The same for the subset of \mgii
    systems with 2~\AA\ $\le$ \ew $<6$~\AA. \emph{Bottom panel:} show
    the \sloii\ as a function of redshift for two \ew\ bins of
    1~\AA\ $\le$ \ew $<2$~\AA\ (\emph{left panel}) and 2~\AA\ $\le$
    \ew $<6$~\AA\ (\emph{right panel}), respectively. The \emph{dashed
      line} show the luminosity surface density evolution of 0.03\ls
    (\emph{left panel}) and 0.1\ls (\emph{right panel}) galaxy as a
    function of redshift.}
\label{fig:loiivsz}
 \end{figure*}

Furthermore, we try to explore the contribution of direct detections
to the stacked spectra. For this, we have selected systems with
1~\AA\ $\le$ \ew $<2$~\AA\ and $0.55 \le z < 0.75$ where we have
sufficient number of systems and the \oii\ emission falls in the
wavelength range free from most crowded telluric emission line region.
Here, if one considers systems with {\rm {\ensuremath{\mathcal{R}}}}
$\ge 0.5$ where the \oii\ emission is clearly detected, the \loii\ is
found to be $(1.88 \pm 0.18) \times 10^{40} \rm ~erg\ s^{-1}$ (i.e,
$\sim 0.047$ \lsoii; where, log~\lsoii\ $=41.6$ at average $z$ of
$\sim 0.65$). However, if we exclude the candidate \oii\ emitters
(i.e., systems with \oii\ emission detected at $\ge 3 \sigma$ level),
which accounts for $\sim$5\% of the systems, we still detect the
\oii\ emission with slightly lower \loii\ of (1.45+0.19) $\times
10^{40} \rm ~erg\ s^{-1}$ (i.e., 0.036\lsoii), albeit consistent
within 1.6$\sigma$ level. Here, it is interesting to ask that what
kind of galaxies do contribute to this \loii. For this, we first
compute the average \loii\ of galaxies by using the \oii\ luminosity
function by \citet{Comparat2016MNRAS.461.1076C} at average $z$ of
0.65, for different lower limits on \loii\ ranging between $ \rm
L_{min} =$ 0.001 - 0.01\lsoii, as:

 \begin{equation}
   \rm \langle L \rangle = \frac {\int_{L_{min}}^{L_{max}} L \phi(L)
  dL} {{\int_{L_{min}}^{L_{max}} \phi(L)dL}} . 
\label{eq:ewvsd} 
\end{equation}

The average \loii\ expected for different values of $\rm L_{min}$ are
listed in column 2 of Table~\ref{tab:lf}. It varies from
0.10\lsoii\ to 0.03\lsoii\ for the $ \rm L_{min}$ ranging from
$0.01$\lsoii\ and $0.001$~\lsoii, respectively. Assuming that the
average \loii\ detected in the stacked spectra after removal of
candidate \oii\ emitters is mainly due to galaxies with \loii\ smaller
than that seen in case of direct detections, we compute the average
\loii\ by restricting ourselves to $\rm L_{max} = 0.23$\lsoii. This
upper limit is set to the average luminosity of lower 5\% systems from
the cumulative distribution of \oii\ luminosity of candidate
\oii\ emitters while considering it as the lowest \loii\ from direct
detections. It is clear from Table ~\ref{tab:lf}, that we recover the
observed \loii\ (i.e., $\sim$ 0.05\lsoii) seen in the stacked spectra
of all the \mgii\ systems when we integrate down to 0.003\lsoii.
Interestingly, the average \loii\ obtained by integrating the
luminosity function over the luminosity range of 0.003\lsoii\ to
0.23\lsoii\ is similar to what we find in the stacked spectra without
direct detections (i.e., $\sim 0.03$\lsoii). This clearly suggest that
strong \mgii absorbers also originate from low luminosity galaxies at
small impact parameters. Note the above estimate is based on the
assumption that galaxies are like point sources and without
considering fibre losses. This should be considered more as an
indicative result. The actual calculations should include size of
galaxies, their orientations and fibre losses into account. We
postpone this for a future work. In addition, the importance of low
mass galaxies contributing to high \ew\ absorbers can be probed
through clustering analysis.

 \begin{table}
 \centering
 {\scriptsize
 \caption{The average \oii\ luminosity in the stacked spectra.}
 \label{tab:lf}
 \begin{tabular}{@{} r c c   c @{}}
 \hline  \hline  
 \multicolumn{1}{c}{$\rm L_{min}$}     &\multicolumn{2}{c}{Average \loii\textcolor{blue}{$^a$}}  \\
 \multicolumn{1}{c}{($\times$ \lsoii)}  &\multicolumn{1}{c}{$\rm \langle L \rangle$ } & \multicolumn{1}{c}{$\rm \langle L^r \rangle$ = $ \frac {\rm \int_{L_{min}}^{0.23L*} L
   \phi(L) dL}{{\rm \int_{L_{min}}^{\infty} \phi(L)dL}}$} \\
\hline

 0.01      &  0.10&  0.042  \\
 0.005     &  0.07&  0.031  \\
 0.003     &  0.05&  0.024  \\
 0.001     &  0.03&  0.015  \\

 \hline                                                                                 
 \end{tabular} 
\\
 } \textcolor{blue}{$^a$}{The upper limits is infinity for the second
   column and 0.23\lsoii\ in the numerator for the third column.}
 \end{table}

 \begin{figure*}
  \centering
    \epsfig{figure=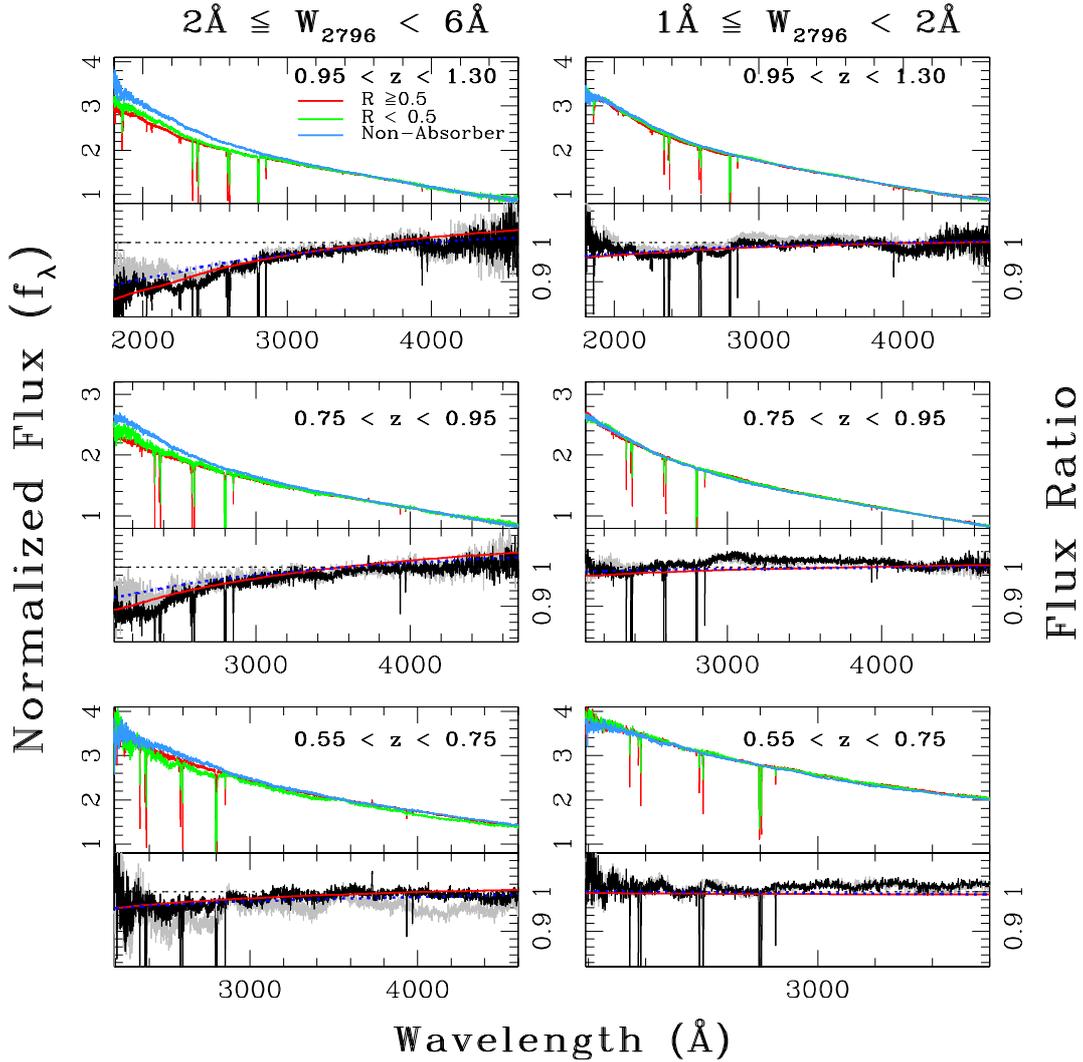,height=14.0cm,width=14.5cm,bbllx=19bp,bblly=147bp,bburx=592bp,bbury=716bp,clip=true}
  \caption{\emph{Left panel:} The geometric mean composite spectra for
    the sub-samples with 2~\AA\ $\le$ \ew\ $<$ 6~\AA\ and three
    redshift bins of $0.55 \le z < 0.75$ (bottom panel), $0.75 \le z <
    0.95$ (middle panel) and $0.95 \le z < 1.3$ (top panel),
    respectively.  The lower part of each panel shows the flux
      ratio of the two spectra for the systems with {\rm
        {\ensuremath{\mathcal{R}}}} $\ge 0.5$ and control sample
      (black) along with the best fit SMC extinction curve
      (\emph{solid red line}). The flux ratio for the systems with
      {\rm {\ensuremath{\mathcal{R}}}} $< 0.5$ and control sample are
      shown in gray along with the best fit SMC extinction curve
      (\emph{dashed blue line}).  \emph{Right panel:} The same as
    left for the subset with 1~\AA\ $\le$ \ew\ $<$ 2~\AA.}
\label{fig:stack_red}
 \end{figure*}

 \emph{Therefore, based on the above discussions we conclude that
   within the impact parameters probed by the SDSS fibre(s) \mgii
   absorbers having higher {\rm {\ensuremath{\mathcal{R}}}} are
   associated with regions having higher \loii. This means, either
   systems with high {\rm {\ensuremath{\mathcal{R}}}} originate from
   regions having high SFR (for a given impact parameter) or have
   smaller impact parameter to the star forming region (for a given
   \ew). Obtaining spatially resolved spectroscopy as well as image
   stacking of these systems \citep{Zibetti2007ApJ...658..161Z} could
   help in discriminating between these two alternatives}.

 \begin{table*}
 \centering
 \begin{minipage}{100mm}
 {\scriptsize
 \caption{The E($B-V$) for various sub-samples.}
 \label{tab:ebmv}
 \begin{tabular}{@{} c c r r r r r  r @{}}
 \hline  \hline 
 \multicolumn{1}{c}{Redshift}    &\multicolumn{4}{c}{E($B-V$)}  \\
 \multicolumn{1}{c}{($z$)}  &\multicolumn{2}{c}{2~\AA\ $\le$ \ew\ $<$6~\AA} &\multicolumn{2}{c}{1~\AA\ $\le$ \ew\ $<$2~\AA} \\
\multicolumn{1}{c}{}  &\multicolumn{1}{c}{\rrr $\ge 0.5$} & \multicolumn{1}{c}{\rrr $< 0.5$} &\multicolumn{1}{c}{\rrr $\ge 0.5$} & \multicolumn{1}{c}{\rrr $< 0.5$}\\
\hline
  
$0.55 \le z < 0.75$  & $0.011\pm0.002$  &  $0.013\pm0.004$  & $-0.001\pm0.002$  &$-0.003\pm0.001$ \\
$0.75 \le z < 0.95$  & $0.027\pm0.001$  &  $0.017\pm0.004$  & $ 0.004\pm0.001$  &$0.002\pm 0.001$ \\
$0.95 \le z < 1.30$  & $0.026\pm0.001$  &  $0.018\pm0.002$  & $ 0.005\pm0.001$  &$0.004\pm0.001$  \\
 \hline                                                                                 
 \end{tabular} 
 }    
 \end{minipage}
 \end{table*}

\subsection{Average dust content}
 \label{sub:red}

 Evidences for the presence of dust in the intervening absorbers are
 commonly seen in the form of continuum reddening. This makes quasar
 absorption systems to be a good tracers of dust content within
 gaseous haloes surrounding galaxies
 \citep{York2006MNRAS.367..945Y,Menard2008MNRAS.385.1053M,Khare2012MNRAS.419.1028K,Menard2012ApJ...754..116M,
   Fukugita2015ApJ...799..195F, Sardane2015MNRAS.452.3192S,
   Murphy2016MNRAS.455.1043M}. Here, we study the dependence of
 average dust content in \mgii\ systems on the \rrr\ parameter, i.e.,
 \rrr $\ge 0.5$ and \rrr $< 0.5$. For this, we have made two subsets
 based on \ew\ with 1~\AA\ $\le$ \ew $<2$~\AA\ and 2~\AA\ $\le$ \ew
 $<6$~\AA. We further divide each subset in to three redshift bins of
 $0.55 \le z < 0.75$, $0.75 \le z < 0.95$ and $0.95 \le z < 1.3$. We
 have generated geometric mean spectra for various sub-samples as well
 as for the control samples of quasars, within $\Delta z = \pm 0.05$
 of \zem and $\Delta r_{\rm mag} = \pm 0.5$ of $r_{\rm mag}$, without
 absorption in their spectra. \par

The stacked spectra for various subsets are shown in
Fig.~\ref{fig:stack_red}. We estimate the reddening, E($B-V$), by
fitting the spectral energy distribution (SED) of the control sample,
reddened by the Small Magellanic Cloud (SMC) extinction curves
\citep{Gordon2003ApJ...594..279G}. The flux ratio of the composite
spectra for the systems with {\rm {\ensuremath{\mathcal{R}}}} $\ge
0.5$ (\emph{black}) and {\rm {\ensuremath{\mathcal{R}}}} $< 0.5$
(\emph{gray}) to the control sample are shown in the lower panel of
Fig.~\ref{fig:stack_red}. The best fit SMC extinction curve for {\rm
  {\ensuremath{\mathcal{R}}}} $\ge 0.5$ and {\rm
  {\ensuremath{\mathcal{R}}}} $< 0.5$ sample are overlaid in
\emph{solid red} line and \emph{dotted} line, respectively. We perform
a bootstrap analysis to measure the uncertainties over each
measurement. For this, we make stacked spectra for 1000 sub-samples by
randomly selecting 70 per cent of the sample and measure the E($B-V$)
by fitting the SED of control sample, reddened by SMC extinction
curves. We consider the standard deviation of E($B-V$) distribution as
1$\sigma$ uncertainty.

In Table~\ref{tab:ebmv}, we have summarized the colour excess,
E($B-V$), for each subset. At first, we confirm that the E($B-V$) is
more towards high \ew\ systems
\citep{Budzynski2011MNRAS.416.1871B,Jiang2011ApJ...732..110J} at any
redshift bin. We also find that the E($B-V$) is higher for the systems
at high redshift \citep[see
  also,][]{Budzynski2011MNRAS.416.1871B,Menard2012ApJ...754..116M}.
For the strong \mgii\ systems (i.e., \ew\ $\ge 1$\AA) in our sample
the E($B-V$) is found to be in the range of -0.001 to 0.027. Using 809
\mgii absorption systems with $1.0 \le$ \zabs\ $\le 1.86$
\citet{York2006MNRAS.367..945Y} have shown that the typical colour
excess, E($B-V$), introduced by these systems ranges from $< 0.001$ to
0.085 \citep[see also,][]{Wild2007MNRAS.374..292W}. Interestingly, for
the subset with 2~\AA\ $\le$ \ew\ $<$6~\AA\ and $0.95 \le z < 1.3$ we
find that \mgii systems with {\rm {\ensuremath{\mathcal{R}}}} $\ge
0.5$ are redder (significant at 3.6$\sigma$ level) than the systems
with {\rm {\ensuremath{\mathcal{R}}}} $< 0.5$. A similar trend is seen
for the subset with 2~\AA\ $\le$ \ew\ $<$6~\AA\ and $0.75 \le z <
0.95$ albeit with a lower significance at 2.4$\sigma$ level. However,
we do not see this trend for any other subsets (see also,
Table~\ref{tab:ebmv}). Note that the typical E($B-V$) of $< 0.02$ has
been inferred using the DLAs \citep{Vladilo2008A&A...478..701V} and
$\sim 0.046$ from the \caii\ absorbers with equivalent width of $\ge
0.7$\AA\ \citep{Sardane2015MNRAS.452.3192S}. Interestingly, the
average E($B-V$) for the sub-sample with 2~\AA\ $\le$ \ew
$<6$~\AA\ and redshift bins of $0.75 \le z < 0.95$ and $0.95 \le z <
1.3$ show the reddening of $\sim$ 0.03 as seen in the extreme-DLAs,
i.e., log~\nhi $\ge 21.7$, by \citet{Noterdaeme2014A&A...566A..24N}.
Moreover, the \hi 21-cm absorbers that also show \rrr\ $\ge 0.5$ tend
to produce more significant reddening in the spectrum of the
background quasars than \mgii\ systems without 21-cm absorbers
\citep{Dutta2017MNRAS.465.4249D}.

\section{Discussion  and conclusions}
\label{lab:dicuss}

We have investigated the effect of fibre size as well as the metal
absorption line ratio (\rratio) on the average luminosity of
\oiiab\ nebular emission from \mgii absorbers (at $0.55 \le z \le
1.3$) in the composite spectra by utilizing quasar spectra obtained
with 3 and 2 arcsec fibres in Sloan Digital Sky Survey. We have found
the following interesting results:

1. We confirm the presence of a strong correlation between
\oii\ luminosity and \ew, in both the data sets. The \sloii\ measured
for SDSS-DR7 is found to be higher than those measured with SDSS-DR12.
This suggests that the fibre effects are not fully taken care of even
when we normalize the luminosity by the projected area. Interestingly,
the difference is found to be largest for the highest \ew\ bin of
$3-6$~\AA. This might be due to the observed large scatter between
\ew\ vs $\rho$ relation which implies that even for the Ultra strong
(\ew $\ge 3$~\AA) \mgii\ systems there is a non negligible probability
of galaxy being outside the fibre (i.e., at large impact parameters).
While discussing the difference between SDSS-DR7 and SDSS-DR12
observations we need to also remember some differences in the
observational strategy adopted. In SDSS-DR7 the fibres are centered on
red whereas in SDSS-DR12 a centering offset is introduced to the
fibres, taking into account atmospheric dispersion, to improve the
flux on blue part. At this stage it is not clear what is the
contribution of this effect to the differences we discuss in this
work. Therefore, it is important to systematically study a sample of
strong Mg~{\sc ii} systems with integral field spectroscopy to map the
extent and nature of star formation associated with Mg~{\sc ii}
systems of different equivalent widths.

2. We also explore the dependence of \loii\ on \rratio. We have found
that the \mgii absorbers with \rrr\ $\ge 0.5$ tend to show higher
\loii\ and \sloii\ (see Fig.~\ref{fig:loiivsew_r}). In fact, the
strong correlation seen between \loii\ vs \ew\ and \loii\ vs $z$ is
mainly driven by systems with strong \feii (i.e. \rrr\ $\ge$ 0.5). For
systems with \rrr\ $< 0.5$ no such trend between \ew\ and \loii\ is
visible. We also show that the fraction of systems with direct
detection increases as a function of \ew\ which gives a hint that the
strong correlation seen in the stacked spectra is possibly a combined
result of the \ew versus $\rho$ anti-correlation and the redshift
dependent fibre losses.

3. Strong dependence of \oii\ luminosity on \rrr\ could mean the
correlation between impact parameter and \rrr\ could be stronger than
that between \ew\ and impact parameter. It will be important to check
this before ascribing any physical connection between \rrr\ and star
formation rate associated with the absorbing galaxy. Unfortunately
\rrr\ values are not available for systems used to define the
correlation between \ew\ and $\rho$
\citep{Nielsen2013ApJ...776..114N}. Therefore, it will be an important
step to explore the correlation between \ew\ and \rrr\ for building a
clear connection between \ew\ and associated star formation rate.

4. We clearly detect the \oii\ emission in the stacked spectra even if
we exclude the candidate \oii\ emitters (systems with nebular emission
detected at $\ge 3 \sigma$ level of significance). This could either
means appreciable contribution from low luminosity galaxies at small
impact parameters or galaxies at larger impact parameters with only
light from outer regions of galaxies contributing to the emission in
the stacked spectra.

5. We confirm the trend of increasing $E(B-V)$ of \mgii absorbers with
increasing \ew\ as well as redshift (see also
Fig.~\ref{fig:stack_red}). Interestingly, for the subset with
2~\AA\ $\le$ \ew\ $<$6~\AA\ and $0.95 \le z < 1.3$ the $E(B-V)$ of
\mgii systems with {\rm {\ensuremath{\mathcal{R}}}} $\ge 0.5$ is found
to be higher than that for the systems with {\rm
  {\ensuremath{\mathcal{R}}}} $< 0.5$ (at 3.6 $\sigma$ level). The
$E(B-V)$ found in {\rm {\ensuremath{\mathcal{R}}}} $> 0.5$ is similar
to the $E(B-V)$ inferred for DLAs. Using the \mgii\ systems searched
for the \hi absorption in STIS survey by
\citet{Rao2006ApJ...636..610R} and \citet{Rao2017MNRAS.471.3428R} we
compute the fraction of strong \mgii systems (\ew $\ge 1$~\AA) having
\rrr\ $\ge 0.5$ being DLAs, i.e., log~\nhi $\ge 20.3$, is $\sim$38\%
which is only $\sim$6 \% in case of \rrr\ $< 0.5$. This fraction
increase to $\sim$60\% if we consider the systems with \ew $\ge
2$~\AA\ and \rrr\ $\ge 0.5$. Therefore, the \mgii absorbers with
\rrr\ $\ge 0.5$ systems may be related to high probability of them
being sub-DLA and DLAs.

\label{lastpage}

\section*{Acknowledgments}
 
 RS, PN, and PPJ acknowledge the support from Indo-French Centre for
 the Promotion of Advance Research (IFCPAR) under project number
 5504$-$2.\par

Funding for the Sloan Digital Sky Survey IV has been provided by the
Alfred P. Sloan Foundation, the U.S. Department of Energy Office of
Science, and the Participating Institutions. SDSS-IV acknowledges
support and resources from the Center for High-Performance Computing
at the University of Utah. The SDSS web site is www.sdss.org.

SDSS-IV is managed by the Astrophysical Research Consortium for the
Participating Institutions of the SDSS Collaboration including the
Brazilian Participation Group, the Carnegie Institution for Science,
Carnegie Mellon University, the Chilean Participation Group, the
French Participation Group, Harvard-Smithsonian Center for
Astrophysics, Instituto de Astrof\'isica de Canarias, The Johns
Hopkins University, Kavli Institute for the Physics and Mathematics of
the Universe (IPMU) / University of Tokyo, Lawrence Berkeley National
Laboratory, Leibniz Institut f\"ur Astrophysik Potsdam (AIP),
Max-Planck-Institut f\"ur Astronomie (MPIA Heidelberg),
Max-Planck-Institut f\"ur Astrophysik (MPA Garching),
Max-Planck-Institut f\"ur Extraterrestrische Physik (MPE), National
Astronomical Observatories of China, New Mexico State University, New
York University, University of Notre Dame, Observat\'ario Nacional /
MCTI, The Ohio State University, Pennsylvania State University,
Shanghai Astronomical Observatory, United Kingdom Participation Group,
Universidad Nacional Aut\'onoma de M\'exico, University of Arizona,
University of Colorado Boulder, University of Oxford, University of
Portsmouth, University of Utah, University of Virginia, University of
Washington, University of Wisconsin, Vanderbilt University, and Yale
University. \bibliography{references}

 \appendix 
\section{}

\subsection{Properties of SDSS-DR12 \mgii systems:} 
\label{sec:apd1}
\begin{figure}
  \centering
  \epsfig{figure=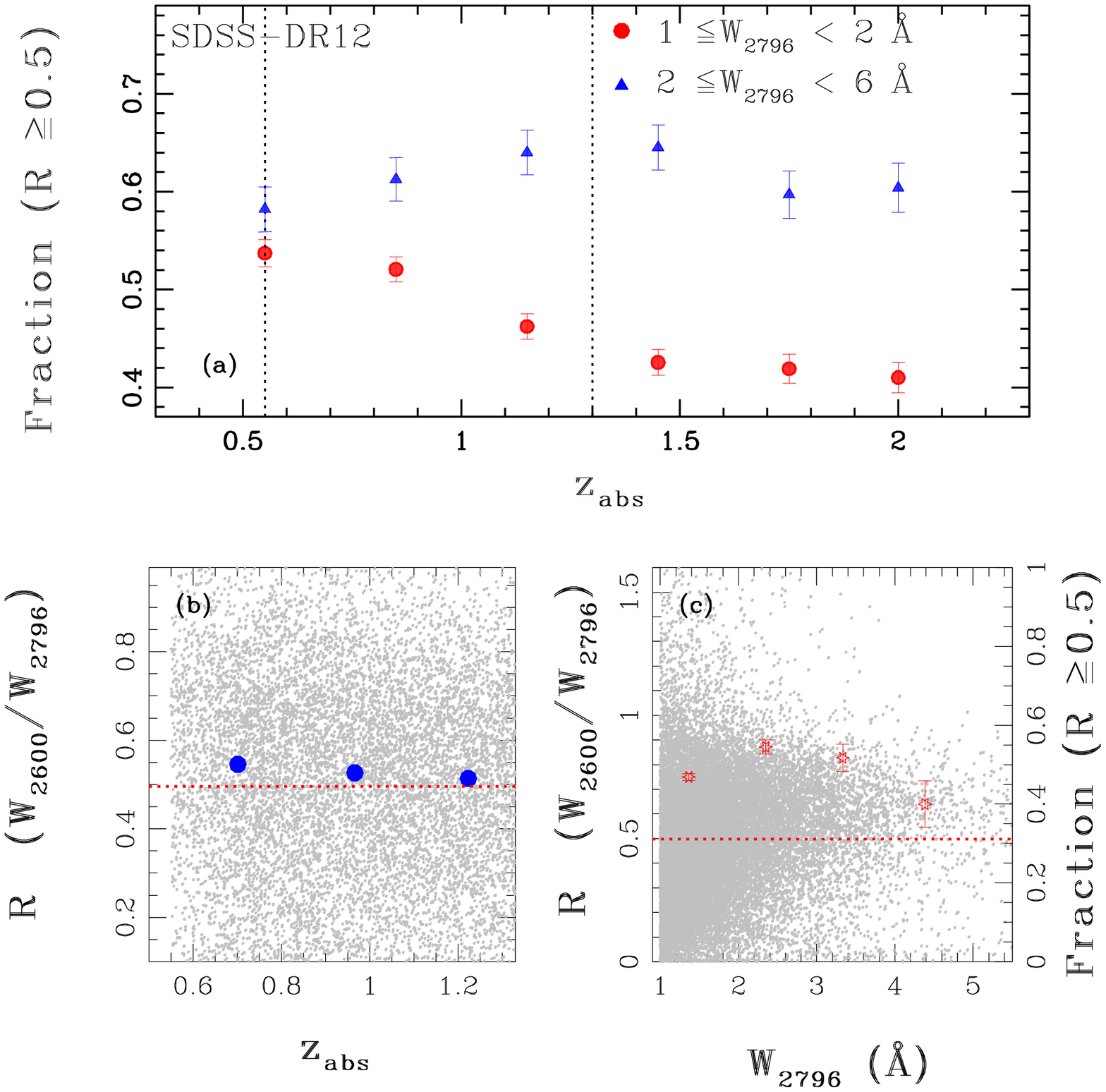,height=9.0cm,width=9.0cm,angle=0}
  \caption{Same as Fig.~\ref{fig:fraction} for \mgii\ systems from
    SDSS-DR12.}
\label{fig:fraction_dr12}
 \end{figure}

In Fig.~\ref{fig:fraction_dr12}, we show the fraction of \mgii systems
with $R \ge 0.5$ as a function of redshift versus $z$ for two
different \ew\ ranges for SDSS-DR12 dataset. We see the similar trend
of \rrr\ as a function of redshift and equivalent width for the
\mgii\ systems detected in SDSS-DR7 data-set.

\subsection{Average  \oii\ emission: Dependence on continuum fit:}
\label{sec:apd2}
\begin{figure*}
  \centering
  \epsfig{figure=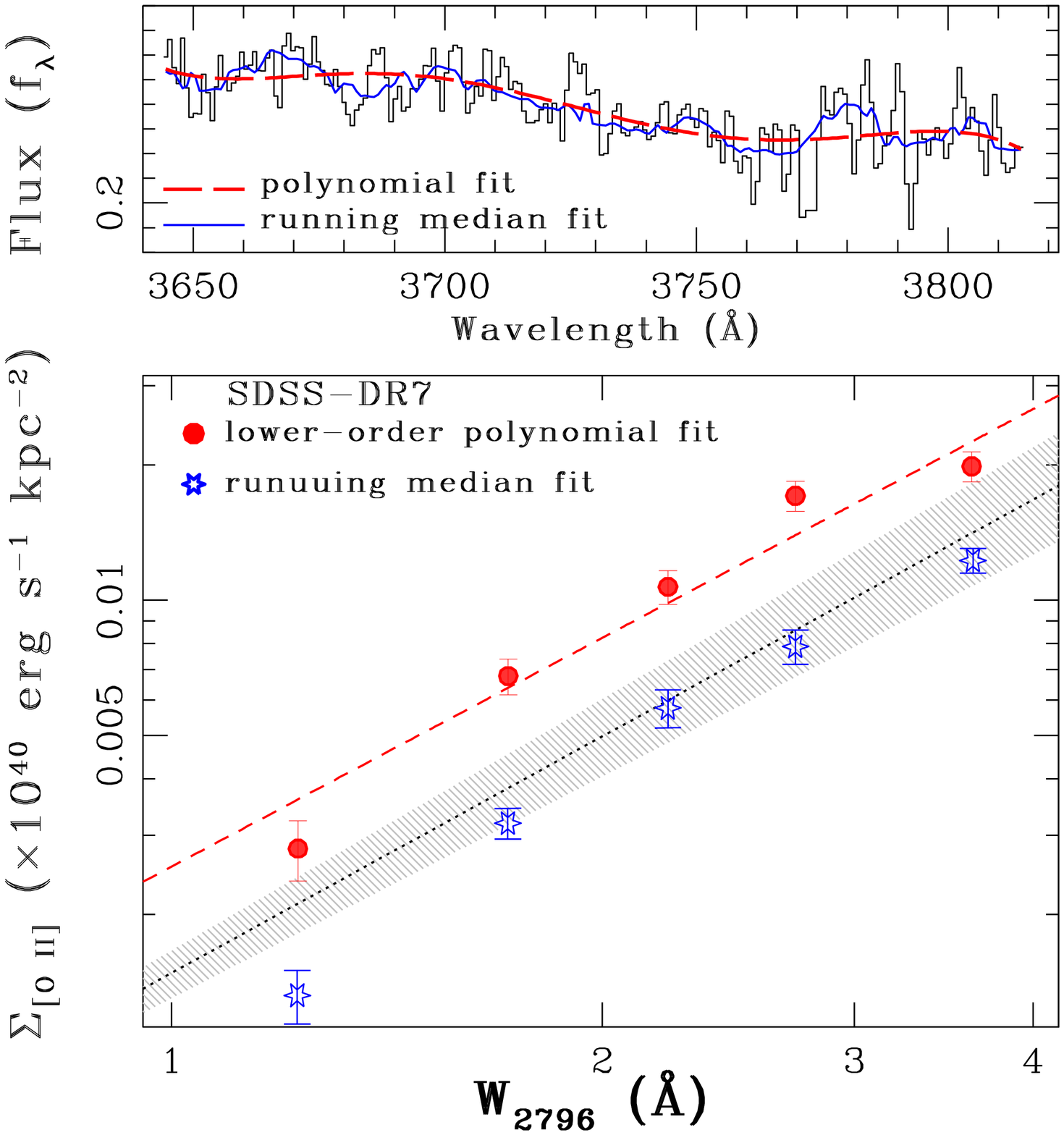,height=13.5cm,width=13.5cm,angle=0}
  \caption{\emph{Lower panel:} The \oii\ luminosity surface density
    (\sloii) as a function of \ew\ for SDSS-DR7 for the stacked
    spectra with continnum modelled as lower order polynomial
    (\emph{red circles}) and the continnum modelled as iterative
    running median of sizes ranging from 500 to 15 pixels (\emph{blue
      stars}). \emph{Upper panel:} The spectral chunk around
    \oii\ nebular emission along with the modelled continnum fit using
    a low order polynoimial (\emph{dashed line}). The continuum fit
    for entire spectra using iterative running median of sizes ranging
    from 500 to 15 pixels is shown in \emph{solid line}.}
\label{fig:loiivsew_menard}
 \end{figure*}

Here, we show that the significant difference in the \sloii\ between
our measurement with \citet{Menard2011MNRAS.417..801M}, as shown in
Fig.~\ref{fig:loiivsew}, is mainly due to the difference in the
modelled continuum. In the upper panel of
Fig.~\ref{fig:loiivsew_menard}, we show the continuum fits for
SDSSJ010226.76+140740.5 modelled by a low order (typically a third
order) polynomial (\emph{dashed line}) and also modelled by an
iterative running median of sizes ranging from 500 to 15 pixels
(\emph{solid line}) as used by \citet{Menard2011MNRAS.417..801M}. It
is apparent from figure that the running median continuum smooths all
the small scale fluctuation and resulting in smaller \oii\ emission.
In the lower panel of Fig.~\ref{fig:loiivsew_menard}, we compare the
\sloii\ as a function \ew\ from the above two methods of continuum
modelling. The \sloii\ measured from the stacked spectra generated
using a local continuum (\emph{circle}) is higher than the one
obtained from running median (\emph{stars}). It is clear from the
figure that the \sloii\ measurement from the continuum fit procedure
of \citet{Menard2011MNRAS.417..801M} are consistent with their best
fit model.

\end{document}